\documentclass{PoS}
\usepackage{slashed}
\usepackage{amsmath}
\usepackage{bbm}
\title{Pion photo- and electroproduction and the chiral MAID interface}

\ShortTitle{Chiral MAID}

\author{Marius Hilt\\
PRISMA Cluster of Excellence, Institut f\"ur Kernphysik,
Johannes Gutenberg-Universit\"at Mainz, D-55099 Mainz, Germany}
\author{Bj\"orn C.~Lehnhart\\
PRISMA Cluster of Excellence, Institut f\"ur Kernphysik,
Johannes Gutenberg-Universit\"at Mainz, D-55099 Mainz, Germany}
\author{\speaker{Stefan Scherer}\\
PRISMA Cluster of Excellence, Institut f\"ur Kernphysik,
Johannes Gutenberg-Universit\"at Mainz, D-55099 Mainz,Germany\\
E-mail: \email{scherer@kph.uni-mainz.de}}
\author{Lothar Tiator\\
PRISMA Cluster of Excellence, Institut f\"ur Kernphysik,
Johannes Gutenberg-Universit\"at Mainz, D-55099 Mainz, Germany}

\abstract{
   We discuss the extended on-mass-shell scheme for
manifestly Lorentz-invariant baryon chiral perturbation theory.
   We present a calculation of pion photo- and electroproduction
up to and including order $q^4$.
   The low-energy constants have been fixed by fitting experimental
data in all available reaction channels.
   Our results can be accessed via a web interface, the so-called chiral MAID
(http://www.kph.uni-mainz.de/MAID/chiralmaid/).
   We explain how our program works and how it can be used for further analysis.}

\FullConference{The 8th International Workshop on Chiral Dynamics, CD2015 ***\\
        29 June 2015 - 03 July 2015\\
        Pisa, Italy}

\begin{document}

\section{Introduction}
\label{sec:introduction}
   In the middle of the 1980s, renewed interest in neutral pion photoproduction at threshold was triggered
by experimental data from Saclay and Mainz~\cite{Mazzucato:1986dz,Beck:1990da}, which indicated a serious
disagreement with the predictions for the $s$-wave electric dipole amplitude $E_{0+}$ based on current algebra
and PCAC \cite{DeBaenst:1971hp,Vainshtein:1972ih,Scherer:1991cy}.
   This discrepancy was explained with the aid of chiral perturbation theory (ChPT)~\cite{Bernard:1991rt}.
   Pion loops, which are beyond the current-algebra framework,
generate infrared singularities in the scattering amplitude which
then modify the predicted low-energy expansion of $E_{0+}$
(see also Ref.\ \cite{Davidson:1993et}).
   Subsequently, several experiments investigating pion photo- and
electroproduction in the threshold region were performed at Mainz,
MIT-Bates, NIKHEF, Saskatoon and TRIUMF, and on the theoretical side, all of
the different reaction channels of pion photo- and electroproduction
near threshold were extensively investigated by Bernard, Kaiser, and
Mei{\ss}ner within the framework of heavy-baryon chiral perturbation
theory (HBChPT) \cite{Bernard:1992qa}.
   For a complete list of references, see Ref.~\cite{Hilt:2013coa}.
   In the beginning, the manifestly Lorentz-invariant or relativistic formulation of ChPT (RChPT) was
abandoned, as it seemingly had a problem with respect to power counting when loops containing internal nucleon
lines come into play.
   In the meantime, the development of the infrared regularization (IR) scheme \cite{Becher:1999he}
and the extended on-mass-shell (EOMS) scheme \cite{Gegelia:1999gf,Fuchs:2003qc} offered a solution to
the power-counting problem, and RChPT became popular again.

   Here, we give a short introduction to the EOMS scheme and
present its application to a calculation of pion photo- and electroproduction
up to and including order $q^4$ [${\cal O}(q^4)$].
   We present the so-called chiral MAID ($\chi$MAID) \cite{website}
which provides the numerical results of these calculations.

\section{Renormalization and power counting}
   Chiral perturbation theory is the effective field theory of QCD in the low-energy regime \cite{Weinberg:1978kz,Gasser:1983yg,Gasser:1987rb}
(for an introduction, see Refs.~\cite{Scherer:2002tk,Scherer:2012zzd}).
   The prerequisite for an effective field theory program
is (a) a knowledge of the most general effective Lagrangian and (b) an expansion scheme for observables in terms
of a consistent power counting method.

\subsection{Effective Lagrangian and power counting}

  The effective Lagrangian relevant to the one-nucleon sector
consists of the sum of the purely mesonic and $\pi N$ Lagrangians, respectively,
\begin{displaymath}
{\cal L}_{\rm eff}={\cal L}_{\pi}+{\cal L}_{\pi N}={\cal L}_\pi^{(2)}+{\cal L}_\pi^{(4)}+\ldots+{\cal L}_{\pi
N}^{(1)}+{\cal L}_{\pi N}^{(2)}+\ldots,
\end{displaymath}
which are organized in a derivative and quark-mass expansion \cite{Weinberg:1978kz,Gasser:1983yg,Gasser:1987rb}.
   For example, the lowest-order basic Lagrangian ${\cal L}_{\pi N}^{(1)}$,
already expressed in terms of renormalized parameters and fields, is given by
\begin{equation}
\label{LpiN1} {\cal L}_{\pi N}^{(1)}=\bar \Psi \left( i\gamma_\mu
\partial^\mu - m\right) \Psi
-\frac{1}{2}\frac{\texttt{g}_A}{F} \bar \Psi \gamma_\mu \gamma_5 \tau^a \partial^\mu \pi^a \Psi +\ldots,
\end{equation}
where $m$, $\texttt{g}_A$, and $F$ denote the chiral limit of the physical nucleon mass, the axial-vector
coupling constant, and the pion-decay constant, respectively.
   The ellipsis refers to terms containing external fields and
higher powers of the pion fields.
   When studying higher orders in perturbation theory, one encounters
ultraviolet divergences.
   As a preliminary step, the loop integrals are regularized,
typically by means of dimensional regularization.
   For example, the simplest dimensionally regularized integral
relevant to ChPT is given by \cite{Scherer:2002tk}
\begin{eqnarray*}
I(M^2,\mu^2,n)&=&\mu^{4-n}\int\frac{\mbox{d}^nk}{(2\pi)^n}\frac{i}{k^2-M^2+i0^+} =\frac{M^2}{16\pi^2}\left[
R+2\ln\left(\frac{M}{\mu}\right)\right]+O(n-4),
\end{eqnarray*}
where $R=\frac{2}{n-4}-[\mbox{ln}(4\pi)+\Gamma'(1)]-1$ approaches infinity as $n\to 4$.
   The 't Hooft parameter $\mu$ is responsible for the fact that the integral has
the same dimension for arbitrary $n$.
   In the process of renormalization the counter terms are adjusted such that
they absorb all the ultraviolet divergences occurring in the calculation of loop diagrams \cite{Collins:xc}.
   This will be possible, because we include in the Lagrangian all
of the infinite number of interactions allowed by symmetries \cite{Weinberg:1995mt}.
   At the end the regularization is removed by taking the limit $n\to 4$.

   Moreover, when renormalizing, we still have the freedom of choosing a renormalization
prescription.
   In this context the finite pieces of the renormalized couplings  will be adjusted such that
renormalized diagrams satisfy the following power counting:
   a loop integration in $n$ dimensions counts as $q^n$,
pion and nucleon propagators count as $q^{-2}$ and $q^{-1}$, respectively, vertices derived from ${\cal
L}_{\pi}^{(2k)}$ and ${\cal L}_{\pi N}^{(k)}$ count as $q^{2k}$ and $q^k$, respectively.
   Here, $q$ collectively stands for a small quantity such as the pion
   mass, small external four-momenta of the pion, and small external
three-momenta of the nucleon.
   The power counting does not uniquely fix the renormalization scheme,
i.e., there are different renormalization schemes such as the IR \cite{Becher:1999he}
and EOMS \cite{Gegelia:1999gf,Fuchs:2003qc} schemes, leading to the above
specified power counting.

\subsection{Example: One-loop contribution to the nucleon mass}

\begin{figure}[t]
\begin{center}
\includegraphics[width=0.5\textwidth]{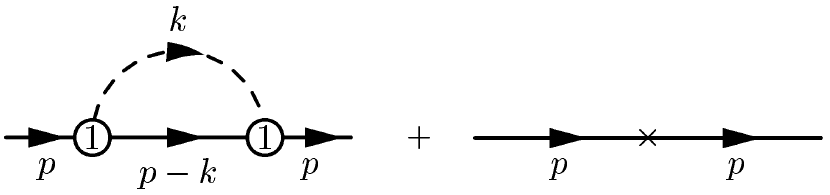}
\caption{\label{4:2:3:ren_diag} Renormalized one-loop self-energy
diagram. The number 1 in the interaction blobs refers to ${\cal L}_{\pi N}^{(1)}$.
The cross generically denotes counter-term contributions.
}
\end{center}
\end{figure}

   In the mesonic sector, the combination of dimensional regularization and
the modified minimal subtraction scheme $\widetilde{\mbox{MS}}$ leads to a straightforward correspondence between
the chiral and loop expansions \cite{Gasser:1983yg}.
   By discussing the one-loop contribution of Fig.~\ref{4:2:3:ren_diag}
to the nucleon self energy, we will see that this correspondence, at first sight, seems to be lost in the
baryonic sector.
   According to the power counting specified above, after renormalization,
we would like to have the order $D=n\cdot 1-2\cdot 1-1\cdot 1+1\cdot 2=n-1.$
   An explicit calculation yields \cite{Scherer:2012zzd}
\begin{displaymath}
\Sigma_{\rm loop}= -\frac{3 \texttt{g}_{A}^2}{4 F^2}\left\{
(\slashed{p}+m)I_N+M^2(\slashed{p}+m)I_{N\pi}-\frac{(p^2-m^2)\slashed{p}}{2p^2}[
(p^2-m^2+M^2)I_{N\pi}+I_N-I_\pi]\right\},
\end{displaymath}
where $M^2=2B\hat m$ is the lowest-order expression for the squared pion mass in terms of the low-energy coupling
constant $B$ and the average light-quark mass $\hat m$ \cite{Gasser:1983yg}.
   The relevant loop integrals are defined as
\begin{eqnarray}
\label{IpiIN}
I_\pi&=& \mu^{4-n}\int\frac{\mbox{d}^nk}{(2\pi)^n}\frac{i}{k^2-M^2+i0^+},\quad\quad
I_N=\mu^{4-n}\int\frac{\mbox{d}^nk}{(2\pi)^n}\frac{i}{k^2-m^2+i0^+},\\
\label{INpi} I_{N\pi}&=&\mu^{4-n}\int\frac{\mbox{d}^nk}{(2\pi)^n}
\frac{i}{[(k-p)^2-m^2+i0^+]}\frac{1}{k^2-M^2+i0^+}.
\end{eqnarray}
   The application of the $\widetilde{\rm MS}$ renormalization scheme of ChPT
\cite{Gasser:1983yg,Gasser:1987rb}---indicated by ``r''---yields
\begin{displaymath}
\Sigma_{\rm loop}^r=-\frac{3 \texttt{g}_{Ar}^2}{4 F^2_r} \left[M^2(\slashed{p}+m)I_{N\pi}^r+\ldots\right].
\end{displaymath}
   The expansion of $I_{N\pi}^r$ is given by
\begin{displaymath}
\label{4:2:3:INpExp}
    I_{N\pi}^r=\frac{1}{16\pi^2}\left(-1+\frac{\pi M}{m}+\ldots\right),
\end{displaymath}
resulting in $\Sigma_{\rm loop}^r={\cal O}(q^2)$.
   In other words, the $\widetilde{\rm MS}$-renormalized result does not
produce the desired low-energy behavior which, for a long time, was interpreted as the absence of a systematic
power counting in the relativistic formulation of ChPT.

   The expression for the nucleon mass $m_N$ is obtained by solving
the equation
\begin{displaymath}
\label{4:2:3:MassDef} m_N-m-\Sigma(m_N)=0,
\end{displaymath}
from which we obtain for the nucleon mass in the $\widetilde{\rm MS}$ scheme \cite{Gasser:1987rb},
\begin{equation}
\label{4:2:3:MassMStilde}
    m_N=m-4c_{1r}M^2+
    \frac{3\texttt{g}_{Ar}^2M^2}{32\pi^2F^2_r}m
    -\frac{3\texttt{g}_{Ar}^2M^3}{32\pi F^2_r}.
\end{equation}
   At ${\cal O}(q^2)$, Eq.~(\ref{4:2:3:MassMStilde}) contains, besides the undesired loop
contribution proportional to $M^2$, the tree-level contribution $-4c_{1r}M^2$ from the next-to-leading-order
Lagrangian ${\cal L}_{\pi N}^{(2)}$.

   The solution to the power-counting problem is the observation
that the term violating the power counting, namely, the third on the right-hand side of
Eq.~(\ref{4:2:3:MassMStilde}), is \emph{analytic} in the quark mass and can thus be absorbed in counter terms.
   In addition to the $\widetilde{\rm MS}$ scheme we have to perform an additional
{\em finite} renormalization.
   For that purpose we rewrite
\begin{equation}
\label{4:2:3:cRenorm}
    c_{1r}=c_1+\delta c_1,\quad \delta c_1 =\frac{3 m {\texttt g}_A^2}{128 \pi^2 F^2_r}+\ldots
\end{equation}
in Eq.~(\ref{4:2:3:MassMStilde}) which then gives the final result for the nucleon mass at ${\cal O}(q^3)$:
\begin{equation}
\label{4:2:3:MassFinal}
    m_N=m-4c_{1}M^2
    -\frac{3\texttt{g}_{A}^2M^3}{32\pi F^2}.
\end{equation}
   We have thus seen that the validity of a power-counting scheme is intimately
connected with a suitable renormalization condition.
   In the case of the nucleon mass, the $\widetilde{\rm MS}$ scheme alone does not
suffice to bring about a consistent power counting.

\subsection{Extended on-mass-shell scheme}
   We illustrate the underlying ideas of the EOMS scheme in terms of a typical one-loop integral
in the chiral limit,
\begin{displaymath}
H(p^2,m^2;n)= -i\int \frac{{\mbox d}^n k}{(2\pi)^n} \frac{1}{[(k-p)^2-m^2+i0^+][k^2+i0^+]},
\end{displaymath}
where $\Delta=(p^2-m^2)/m^2={\cal O}(q)$ is a small quantity.
   Applying the dimensional counting analysis of
Ref.~\cite{Gegelia:1994zz}, the result of the integration is of the form
\begin{displaymath}
H\sim F(n,\Delta)+\Delta^{n-3}G(n,\Delta),
\end{displaymath}
where $F$ and $G$ are hypergeometric functions which are analytic for $|\Delta|<1$ for any $n$.
   The central idea of the EOMS scheme \cite{Gegelia:1999gf,Fuchs:2003qc} consists of
subtracting those terms which violate the power counting as $n\to 4$.
   Since the terms violating the power counting are analytic in small
quantities, they can be absorbed by counter-term contributions.
   In the present case, we want the renormalized integral to be of
the order $D=n-1-2=n-3$.
   To that end one first expands the integrand in
small quantities and subtracts those integrated terms whose order is smaller than suggested by the power
counting.
   The corresponding subtraction term reads \cite{Fuchs:2003qc}
\begin{displaymath}
H^{\rm subtr}=-i\int \frac{\mbox{d}^n k}{(2\pi)^n}\left. \frac{1}{[k^2-2p\cdot k +i0^+][k^2+i0^+]}\right|_{p^2=m^2}
=\frac{m^{n-4}}{(4\pi)^\frac{n}{2}}\frac{\Gamma\left(2-\frac{n}{2}\right)}{n-3},
\end{displaymath}
and the renormalized integral is written as
\begin{displaymath}
H^R=H-H^{\rm subtr}=\frac{m^{n-4}}{(4\pi)^{\frac{n}{2}}}\left[-\Delta\ln(-\Delta)+(-\Delta)^2\ln(-\Delta)+\ldots\right]
={\cal O}(q)\quad\text{as $n\to 4$.}
\end{displaymath}
   The EOMS scheme can be straightforwardly extended to include other degrees of freedom,
such as vector mesons \cite{Fuchs:2003sh} or the $\Delta(1232)$
resonance \cite{Hacker:2005fh}, and it can be applied to multi-loop diagrams \cite{Schindler:2003je}.
   Moreover, the infrared regularization of Becher and Leutwyler \cite{Becher:1999he} has been
reformulated in a form analogous to the EOMS renormalization \cite{Schindler:2003xv}.
   Applications include the calculation of the electromagnetic form factors of the nucleon
\cite{Fuchs:2003ir,Schindler:2005ke,Bauer:2012pv}, the axial form factor \cite{Schindler:2006it},
pion-nucleon scattering \cite{Alarcon:2011zs,Alarcon:2012kn,Chen:2012nx},
and a two-loop analysis of the nucleon mass \cite{Schindler:2006ha,Schindler:2007dr}.
   For a review of the three-flavor sector, see Ref.~\cite{Geng:2013xn}.

   The inclusion of virtual vector mesons generates an improved description of
the electromagnetic form factors \cite{Schindler:2005ke,Bauer:2012pv}, for which
ordinary chiral perturbation theory does not produce sufficient curvature.
   However, one would also like to describe the properties of hadronic resonances such as their masses
and widths as well as their electromagnetic properties.
   In this context, one needs to set up a power counting valid in the momentum region near the complex
pole of a resonance.
   An extension of the chiral effective field theory program in this direction was proposed
in Ref.~\cite{Djukanovic:2009zn}, in which the power-counting problem
was addressed by applying the complex-mass scheme (CMS)
\cite{Stuart:1990,Denner:1999gp} to the effective field theory.
   For recent applications of the CMS to hadronic resonances, see
Refs.~\cite{Djukanovic:2009gt,Bauer:2012at,Djukanovic:2013mka,Bauer:2014cqa,Djukanovic:2014rua,Epelbaum:2015vea}.

\section{Pion photo- and electroproduction}
\label{sec:electroproduction}

\subsection{Invariant amplitude and cross section}
\begin{figure}[ht]
\begin{center}
\includegraphics[width=0.5\textwidth]{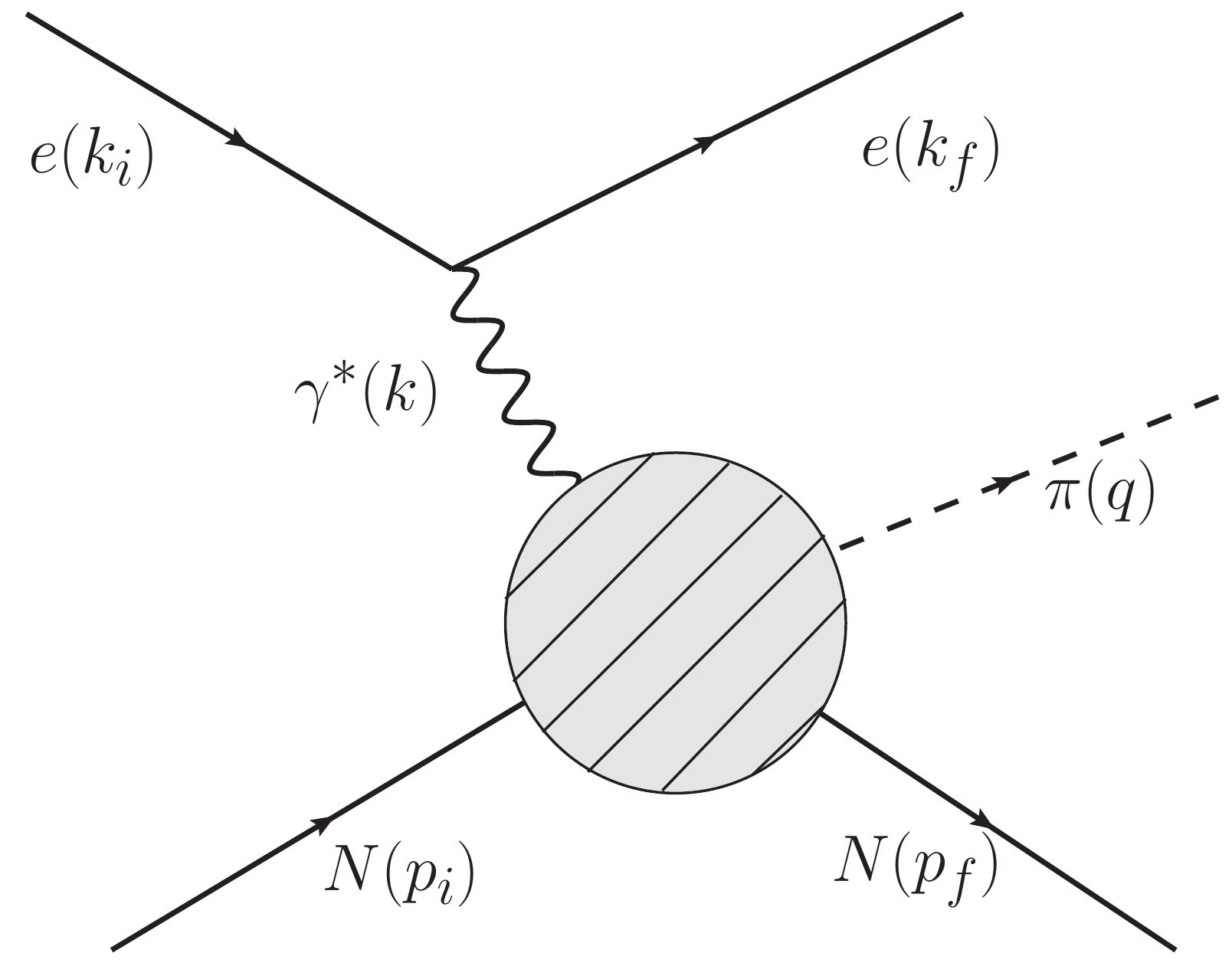}
\caption{\label{fig:one-photon-exchange-approximation} Pion electroproduction in the one-photon-exchange approximation.
The momenta of the incoming and outgoing nucleons are $p_i$ and $p_f$, respectively.
    The momentum of the incoming/outgoing electron is $k_i$/$k_f$, where $k=k_i-k_f$ represents the
    momentum of the single exchanged virtual photon.
    The momentum of the pion is labeled $q$.
    In the case of pion photoproduction, the leptonic vertex and the photon propagator are
    replaced by the polarization vector of the real photon.
    The shaded circle represents the full hadronic vertex.
}
\end{center}
\end{figure}

   In the one-photon-exchange approximation (see Fig.~\ref{fig:one-photon-exchange-approximation}), the invariant amplitude
for the reaction $e(k_i,s_i)+N(p_i,S_i)\rightarrow e(k_f,s_f)+N(p_f,S_f)+\pi(q)$ may be written as
\begin{equation}
\label{eq:invariant_amplitude}
{\cal M}=\epsilon_\mu {\cal M}^\mu,\quad \epsilon_\mu=e\frac{\bar{u}(k_f,s_f)\gamma_\mu u(k_i,s_i)}{k^2},\quad k=k_i-k_f,
\end{equation}
where $e>0$ is the elementary charge, $\epsilon_\mu$ denotes the polarization vector of the virtual photon,
and ${\cal M}^\mu$ is the hadronic transition current matrix element:
${\cal M}^\mu=-ie\langle N(p_f,S_f),\pi(q)|J^\mu(0)|N(p_i,S_i)\rangle$.
      Therefore, it is sufficient to consider the process
\begin{equation}´
\gamma^*(k)+N(p_i)\rightarrow N(p_f)+\pi(q),
\end{equation}
where $\gamma^\ast$ refers to a virtual photon.
   The invariant amplitude of pion photoproduction is obtained by replacing the polarization
vector of the virtual photon by the polarization vector of a real photon and taking $k^2=0$.
   Treating the virtual photon as a particle of ``mass'' $k^2=-Q^2$, the Mandelstam variables
$s$, $t$, and $u$ are defined as
\begin{equation}
s=(p_i+k)^2=(p_f+q)^2,\ t=(p_i-p_f)^2=(q-k)^2, \ u=(p_i-q)^2=(p_f-k)^2,\label{eqn:mandelstam}
\end{equation}
and fulfill
\begin{equation}
s+t+u=2m_N^2+M_\pi^2-Q^2,
\end{equation}
where $m_N$  and $M_\pi$ denote the nucleon mass and the pion mass, respectively.
   In the case of photoproduction ($k^2=0$) only two of the Mandelstam variables are independent.
   In the center-of-mass (c.m.) frame (see Fig.~\ref{fig:cm_frame}), the energies of the photon, $k_0$, and the pion, $E_\pi$,
are given by
\begin{equation}
k_0=\frac{W^2-m_N^2-Q^2}{2W},\ E_\pi=\frac{W^2+M_\pi^2-m_N^2}{2W},
\end{equation}
where $W=\sqrt{s}$ is the c.m.~total energy.
   The equivalent real photon laboratory energy $E_\gamma^\textnormal{lab}$ is given by
\begin{equation}
E_\gamma^\textnormal{lab}=\frac{W^2-m_N^2}{2m_N}.
\end{equation}
   The c.m.~scattering angle $\Theta_\pi$ between the pion three-momentum and the $z$-axis,
defined by the incoming (virtual) photon, can be related to the Mandelstam variable $t$ via
\begin{equation}
t=M_\pi^2-2(E_\gamma E_\pi-|\vec{k}||\vec{q}\hspace{0.05cm}|\cos\Theta_\pi).
\end{equation}
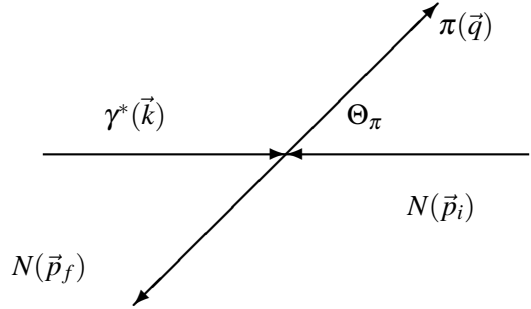
\begin{figure}[ht]
\begin{center}
{
\unitlength0.8cm
\begin{picture}(10,6)
\thicklines
\put(1,3){\vector(1,0){4}}
\put(2,3.5){$\gamma^\ast(\vec{k})$}
\put(9,3){\vector(-1,0){4}}
\put(7,2){$N(\vec{p}_i)$}
\put(5,3){\vector(1,1){2.5}}
\put(7.5,5){$\pi(\vec{q})$}
\put(5,3){\vector(-1,-1){2.5}}
\put(0.5,1){$N(\vec{p}_f)$}
\put(6,3.5){$\Theta_\pi$}
\end{picture}
}
\end{center}
\caption{\label{fig:cm_frame} Kinematics in the c.m.~frame.}
\end{figure}

   Using current conservation, $k_\mu{\cal M}^\mu=0$,
the transition current matrix element may be parametrized in terms of six invariant amplitudes $A_i$,
\begin{equation}
\label{eq:dennery}
\mathcal{M}^\mu=\bar{u}(p_f,S_f)\Bigg(\sum_{i=1}^6 A_i(s,t,u) M_i^\mu\Bigg)u(p_i,S_i).
\end{equation}
   The $M_i^\mu$ are suitable, linearly independent $4\times 4$ matrices,
\begin{align*}
M_1^\mu&=-\frac{i}{2}\gamma_5\left(\gamma^\mu \slashed{k}-\slashed{k}\gamma^\mu\right),\\
M_2^\mu&=2i\gamma_5\left[P^\mu k\cdot\left(q-\frac{1}{2}k\right)
-\left(q^\mu-\frac{1}{2}k^\mu\right) k\cdot P\right],\\
M_3^\mu&=-i\gamma_5\left(\gamma^\mu k\cdot q-\slashed{k}q^\mu\right),\\
M_4^\mu&=-2i\gamma_5\left(\gamma^\mu k\cdot P-\slashed{k}P^\mu\right)-2m_N M_1^\mu,\\
M_5^\mu&=i\gamma_5\left(k^\mu k\cdot q-q^\mu k^2 \right),\\
M_6^\mu&=-i\gamma_5\left(\slashed{k} k^\mu-\gamma^\mu k^2\right),
\end{align*}
where $P=(p_i+p_f)/2$.
   Note that each structure $M_i^\mu$ satisfies $k_\mu M^\mu_i=0$.

   The so-called Chew-Goldberger-Low-Nambu (CGLN) amplitudes $\mathcal{F}_i$ are another
common parametrization \cite{Chew:1957tf,Dennery:1961zz}.
   These amplitudes are defined in the c.m.~frame  via
\begin{equation}
\epsilon_\mu\bar{u}(p_f,S_f)\left(\sum_{i=1}^6A_i M_i^\mu\right) u(p_i,S_i)=\frac{4\pi W}{m_N}\chi_f^\dagger\mathcal{F}\chi_i,
\end{equation}
where $\chi_i$ and $\chi_f$ denote initial and final Pauli spinors.
   Electromagnetic current conservation allows one to work in a gauge where the polarization vector
of the virtual photon has a vanishing time component.
   In terms of the polarization vector of Eq.~(\ref{eq:invariant_amplitude}) this is achieved by introducing
the vector \cite{Amaldi:1979vh}
\begin{equation}
a^\mu=\epsilon^\mu-k^\mu \frac{\epsilon_0}{k_0}=\epsilon^\mu-k^\mu \frac{\vec{k}\cdot\vec{\epsilon}}{k_0^2},
\end{equation}
where use of $k_\mu \epsilon^\mu=0$ has been made.
   Splitting $\vec a$ into a longitudinal and a transversal piece,
\begin{equation}
\begin{split}
\vec a&=\vec a_\parallel+\vec a_\perp,\\
\vec a_\parallel&=\vec{a}\cdot\hat k\,\hat k=\frac{k^2}{k_0^2}\vec\epsilon\cdot\hat k\,\hat k,\\
\vec a_\perp&=\vec a-\vec a_\parallel=\vec \epsilon-\vec\epsilon\cdot\hat k\,\hat k=\vec\epsilon_\perp,
\end{split}
\end{equation}
$\mathcal{F}$ may be written as
\begin{equation}
\mathcal{F}=i \vec{\sigma}\cdot\vec{a}_\perp\mathcal{F}_1
+\vec{\sigma}\cdot\hat{q}\,\vec{\sigma}\cdot\hat{k}\times\vec{a}_\perp\mathcal{F}_2
+i\vec{\sigma}\cdot\hat{k}\,\hat{q}\cdot\vec{a}_\perp\mathcal{F}_3
+i \vec{\sigma}\cdot\hat{q}\,\hat{q}\cdot\vec{a}_\perp\mathcal{F}_4
+i \vec{\sigma}\cdot\hat{k}\,\hat{k}\cdot\vec{a}_\parallel\mathcal{F}_5
+i\vec{\sigma}\cdot\hat{q}\,\hat{k}\cdot\vec{a}_\parallel\mathcal{F}_6,
\label{eq:F}
\end{equation}
where $\hat{q}$ and $\hat{k}$ denote unit vectors in the direction of the pion and the
photon, respectively.
   For the case of pion photoproduction, only the first four terms of Eq.~(\ref{eq:F})
contribute.
   The CGLN amplitudes can be expanded in a multipole series \cite{Chew:1957tf,Dennery:1961zz,Amaldi:1979vh},
\begin{equation}
\begin{split}
\mathcal{F}_1&=\sum_{l=0}^\infty\Big\{\big[lM_{l+}+E_{l+}\big]P'_{l+1}(x)
+\big[(l+1)M_{l-}+E_{l-}\big]P'_{l-1}(x)\Big\},\\
\mathcal{F}_2&=\sum_{l=1}^\infty\Big\{(l+1)M_{l+}+lM_{l-}\Big\}P'_l(x),\\
\mathcal{F}_3&=\sum_{l=1}^\infty\Big\{\big[E_{l+}-M_{l+}\big]P''_{l+1}(x)
+\big[E_{l-}+M_{l-}\big]P''_{l-1}(x)\Big\},\\
\mathcal{F}_4&=\sum_{l=2}^\infty\Big\{M_{l+}-E_{l+}-M_{l-}-E_{l-}\Big\}P''_l(x),\\
\mathcal{F}_5&=\sum_{l=0}^\infty\Big\{(l+1)L_{l+}P'_{l+1}-lL_{l-}P'_l(x)\Big\},\\
\mathcal{F}_6&=\sum_{l=1}^\infty\Big\{lL_{l-}-(l+1)L_{l+}\Big\}P'_l(x),
\end{split}
\label{eq:Fi}
\end{equation}
where $x=\cos\Theta_\pi=\hat{q}\cdot\hat{k}$.
   In Eq.~(\ref{eq:Fi}), $P_l(x)$ is a Legendre polynomial of degree $l$,
$P'_l=dP_l/dx$ and so on, with $l$ denoting the orbital angular momentum
of the pion-nucleon system in the final state.
   The multipoles $E_{l\pm}$, $M_{l\pm}$, and $L_{l\pm}$ are functions of the
c.m.~total energy $W$ and the photon virtuality $Q^2$ and
refer to transversal electric and
magnetic transitions and longitudinal transitions, respectively.
   The subscript $l\pm$ denotes the total angular momentum $j=l\pm1/2$ in the final state.
   In the threshold region, the multipoles $\mathcal{M}_{l\pm}$ ($\mathcal{M}=E,M,L$) are proportional
to $|\vec{q}|^l$.
   To get rid of this purely kinematical dependence, one introduces reduced multipoles
$\overline{\mathcal{M}}_{l\pm}$ via
\begin{equation}
\overline{\mathcal{M}}_{l\pm}=\frac{\mathcal{M}_{l\pm}}{|\vec{q}|^l}.
\end{equation}

   Due to the assumed isospin symmetry, the process involves only three independent isospin structures
for the four physical channels \cite{Chew:1957tf}.
   Any amplitude $M$ for producing a pion with Cartesian isospin index
$a$ can be decomposed as
\begin{equation}
M(\pi^a)=\chi_f^\dagger\left(i\epsilon^{a3b}\tau^bM^{(-)}+\tau^aM^{(0)}+\delta^{a3}M^{(+)}\right)\chi_i,
\qquad a=1,2,3,
\label{eqn:isospin}
\end{equation}
where $\chi_i$ and $\chi_f$ denote the isospinors of the initial and final nucleons, respectively,
and $\tau^a$ are the Pauli matrices.
   The isospin amplitudes corresponding to $A_i$ of Eq.\ (\ref{eq:dennery}) obey a crossing symmetry,
\begin{equation}
\begin{split}
A^{(0,+)}_i&\stackrel{s\leftrightarrow u}{\longrightarrow}\eta_i A^{(0,+)}_i,\\
A^{(-)}_i&\stackrel{s\leftrightarrow u}{\longrightarrow}-\eta_i A^{(-)}_i,
\end{split}
\label{eqn:crossing}
\end{equation}
where $\eta_i=1$ for $i=1,2,4$ and $\eta_i=-1$ for $i=3,5,6$.
   Finally, the physical reaction channels are related to the isospin channels via
\begin{align*}
A_i(\gamma^{(*)}p\rightarrow n\pi^+)&=\sqrt{2}\left(A_i^{(-)}+A_i^{(0)}\right),&
A_i(\gamma^{(*)}n\rightarrow p\pi^-)&=-\sqrt{2}\left(A_i^{(-)}-A_i^{(0)}\right),\\
A_i(\gamma^{(*)}p\rightarrow p\pi^0)&=A_i^{(+)}+A_i^{(0)},
&A_i(\gamma^{(*)}n\rightarrow n\pi^0)&=A_i^{(+)}-A_i^{(0)}.
\end{align*}

    For pion photoproduction with polarized photons from an unpolarized target without recoil
polarization detection, the cross section can be written in the following way with the unpolarized
cross section $\sigma_0$ und the photon beam asymmetry $\Sigma$:
\begin{equation}
\frac{d \sigma}{d \Omega} =  \sigma_0 \left( 1 - P_T \Sigma \cos 2 \varphi \right).
\end{equation}
For $\pi^0$ photoproduction on the proton, both observables are very precisely measured in the
threshold region, allowing for an almost model-independent partial-wave
analysis~\cite{Hornidge:2012ca}.

   For pion electroproduction, in the one-photon-exchange approximation, the differential cross
section can be written as
\begin{equation}
\frac{d\sigma}{d\mathcal{E}_fd\Omega_fd\Omega_\pi^\textnormal{\,c.m.}}
=\Gamma\frac{d\sigma_v}{d\Omega_\pi^\textnormal{\,c.m.}},
\end{equation}
where $\Gamma$ is the virtual photon flux and $d\sigma_v/d\Omega_\pi^\textnormal{\,c.m.}$ is the pion
production cross section for virtual photons.

   For an unpolarized target and without recoil polarization detection, the virtual-photon
differential cross section for pion production can be further decomposed as
\begin{equation}
\frac{d\sigma_v}{d\Omega_\pi}=\frac{d\sigma_T}{d\Omega_\pi} +\epsilon\frac{d\sigma_L}{d\Omega_\pi}
+\sqrt{2\epsilon(1+\epsilon)}\frac{d\sigma_{LT}}{d\Omega_\pi}\cos\Phi_\pi
+\epsilon\frac{d\sigma_{TT}}{d\Omega_\pi}\cos{2\Phi_\pi}
+h\sqrt{2\epsilon(1-\epsilon)}\frac{d\sigma_{LT'}}{d\Omega_\pi}\sin{\Phi_\pi}, \label{eqn:wqparts}
\end{equation}
where it is understood that the variables of the individual virtual-photon cross sections
$d\sigma_T/d\Omega_\pi$ etc.~refer to the c.m.~frame.
   For further details, especially concerning polarization observables, see Ref.~\cite{Drechsel:1992pn}.

\subsection{Evaluation of the invariant amplitude and chiral MAID}
   At ${\cal O}(q^3)$, the invariant amplitude involves 15 tree-level diagrams and 50 one-loop diagrams.
   At ${\cal O}(q^4)$, 20 tree-level diagrams and 85 one-loop diagrams contribute.
   We have calculated the loop contributions numerically, using the computer algebra system MATHEMATICA with
the Feyn Calc \cite{Mertig:1990an} and LoopTools packages \cite{Hahn:2000kx}.
   We have explicitly verified that current conservation and crossing symmetry
are fulfilled analytically for our results.

   At ${\cal O}(q^3)$, four independent LECs exist which are specifically related
to pion photoproduction.
   Two of them enter the isospin ($-$) channel and are, therefore, only relevant
for the production of charged pions.
   Moreover, they contribute differently to the invariant amplitudes $A_i$ of Eq.~(\ref{eq:dennery}).
   The remaining two constants enter the isospin
(+) and (0) channels, respectively, though both in combination with the same Dirac structure.
   Finally, at ${\cal O}(q^3)$ the description of pion electroproduction is a prediction,
because no new parameter (LEC) beyond photoproduction is available at that order.
   At ${\cal O}(q^4)$,  15 additional LECs appear.
   In the case of pion photoproduction, five constants contribute to the isospin (0) channel,
five constants to the isospin (+) channel, and one constant to the isospin ($-$) channel.
   For electroproduction, the (0) and (+) channels each have two more independent LECs.
   We note that the isospin $(-)$ channel, even at ${\cal O}(q^4)$, does not contain any
free LEC specifically related to electroproduction.

\begin{figure}[ht]
\includegraphics[width=\textwidth]{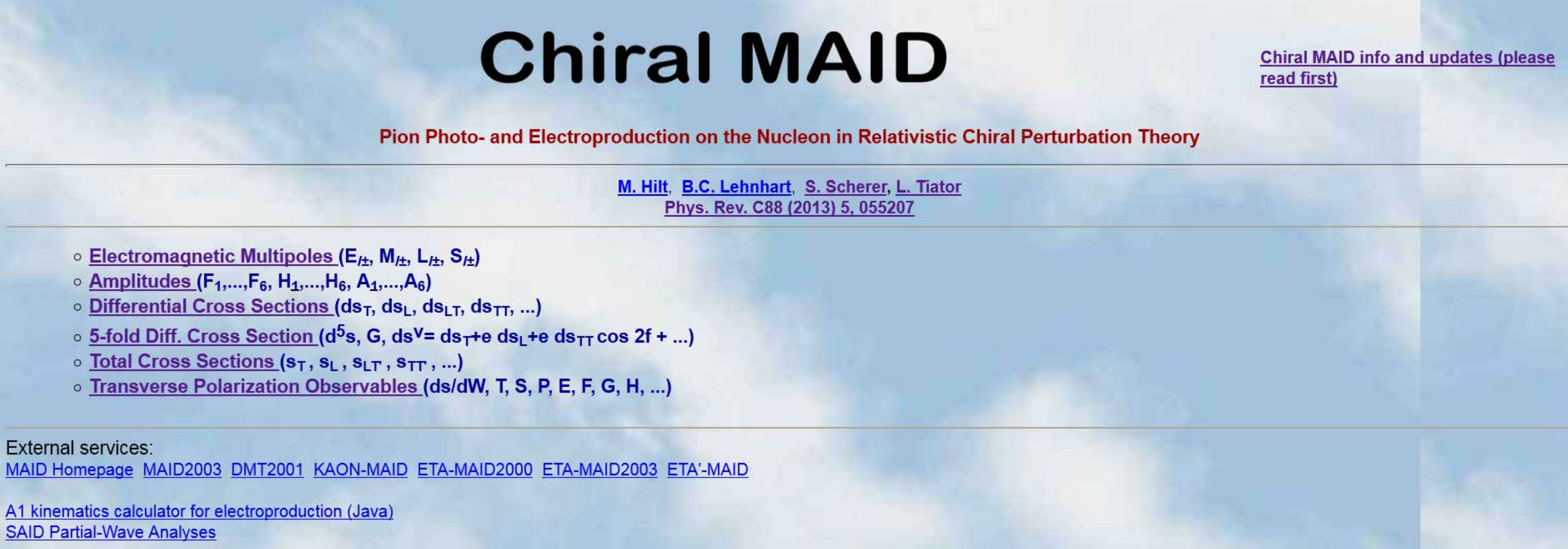}
\caption{\label{fig:chiral_MAID} Chiral MAID homepage [http://www.kph.uni-mainz.de/MAID/chiralmaid/].}
\end{figure}

\begin{figure}[ht]
\begin{center}
\includegraphics[width=\textwidth]{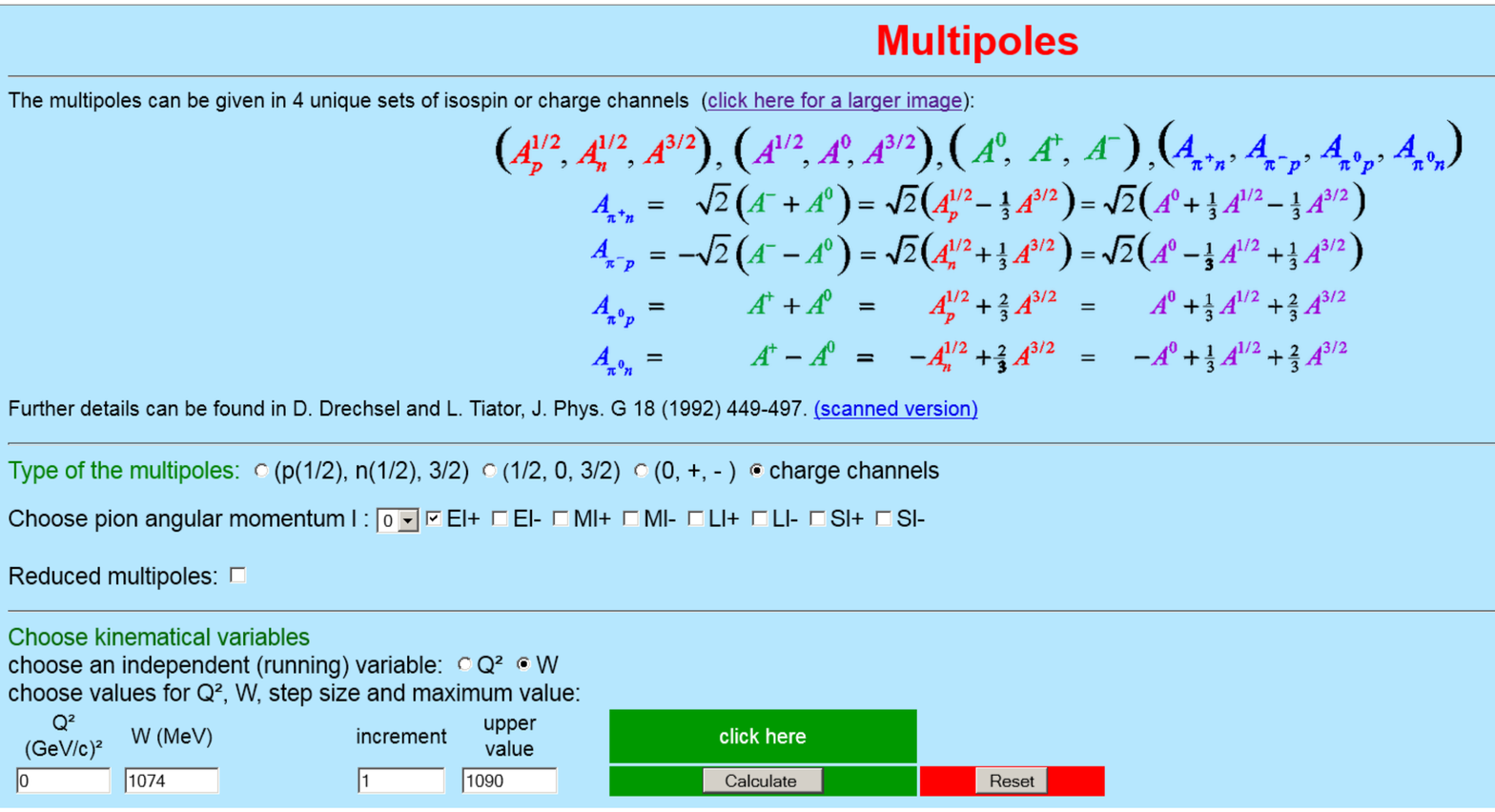}
\caption{\label{fig:Multipoles} Settings to calculate the electric
dipole amplitude $E_{0+}$ for the physical channels.}
\end{center}
\end{figure}

\begin{figure}[ht]
\begin{center}
\includegraphics[width=\textwidth]{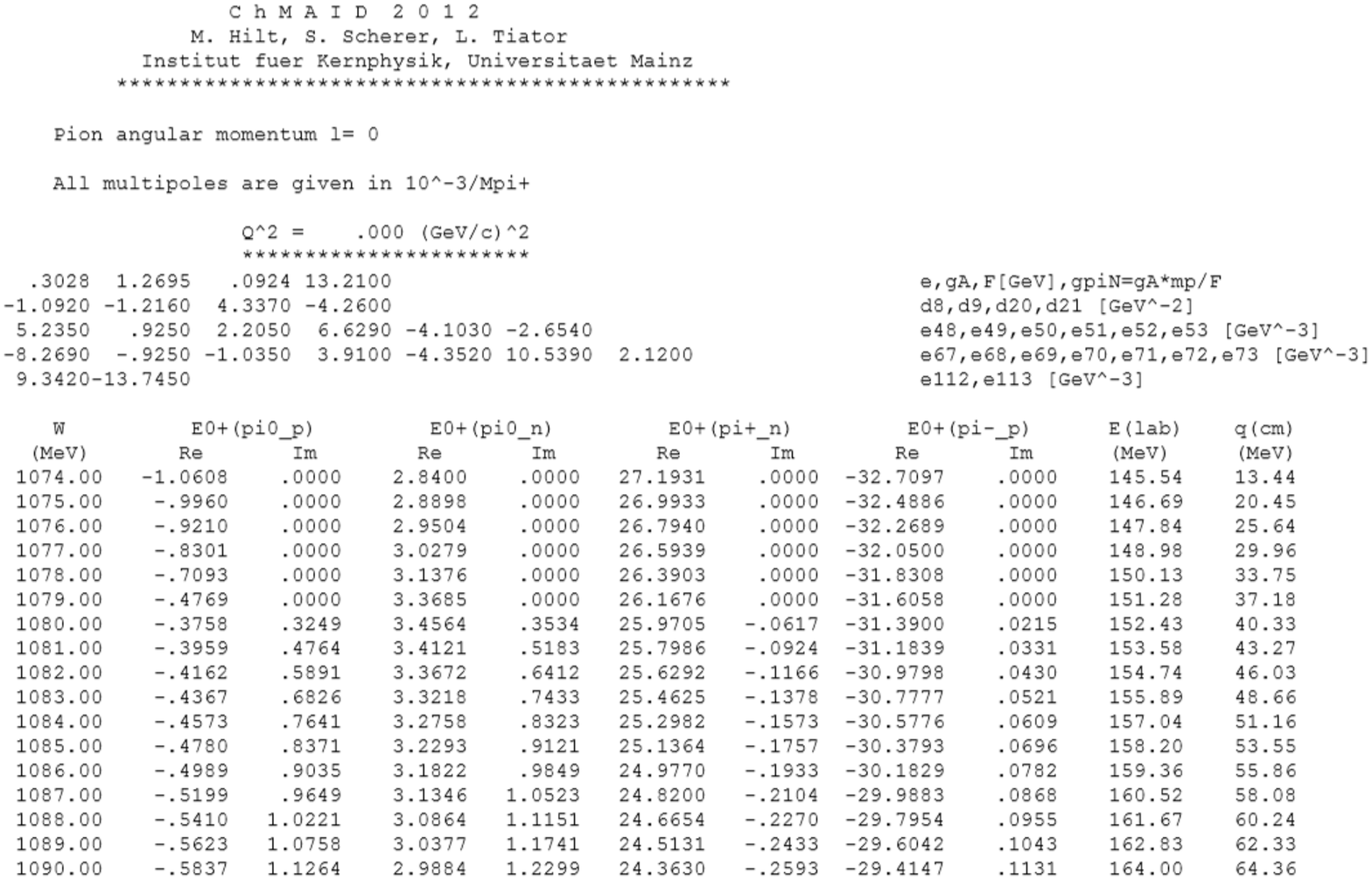}
\caption{\label{fig:Output} Output for the electric
dipole amplitude $E_{0+}$ for the physical channels.}
\end{center}
\end{figure}

\begin{figure}[ht]
\begin{center}
\includegraphics[width=0.8\textwidth]{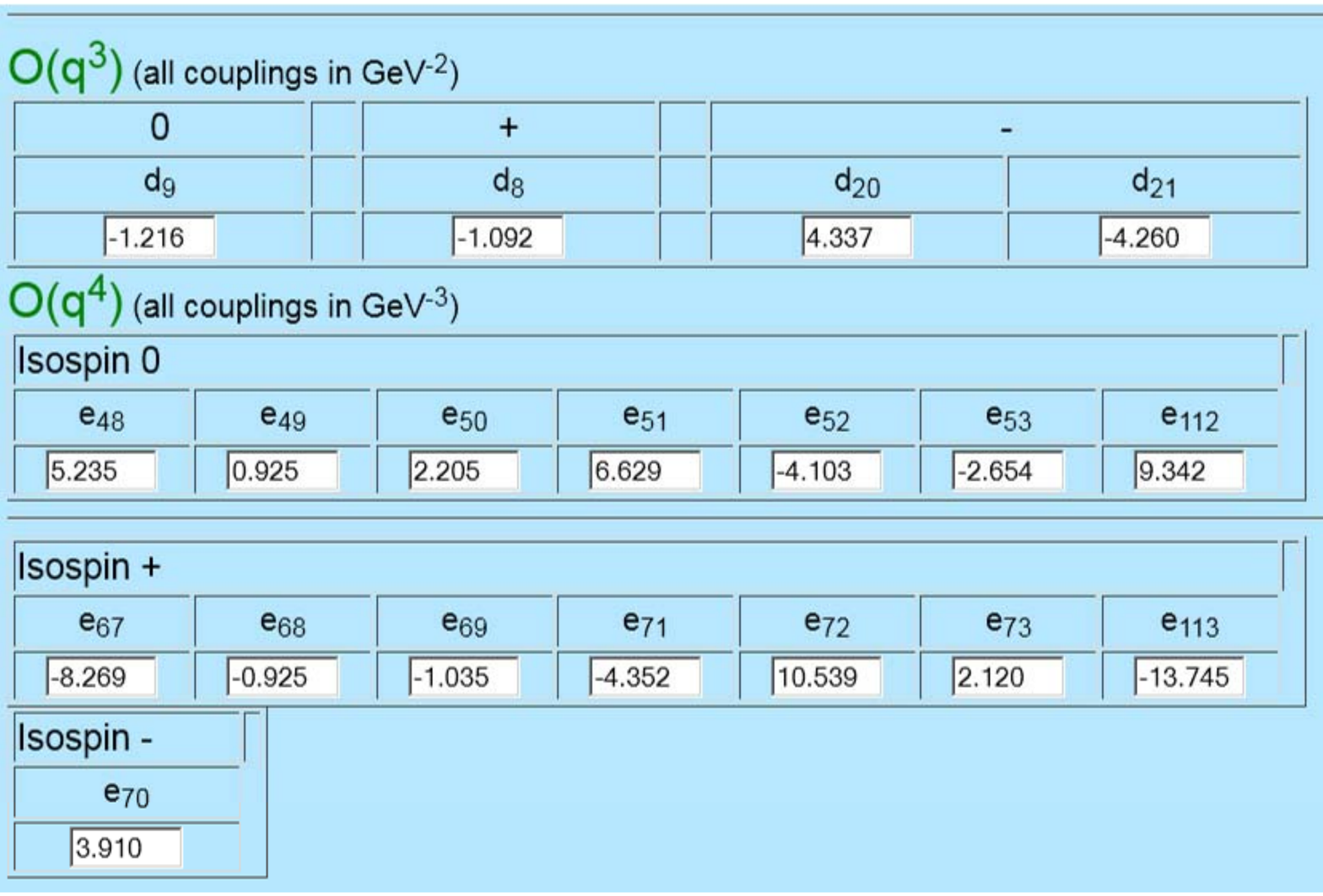}
\caption{\label{fig:Change_of_parameters} The LECs of the contact interactions can be
modified by the user.}
\end{center}
\end{figure}

   Figure \ref{fig:chiral_MAID} shows the homepage of the web interface chiral MAID.
   At the beginning, the program allows one to choose among various quantities to be calculated (multipoles,
different sets of amplitudes, etc.).
   The loop contributions, including their parameters, are fixed and cannot be modified from the outside.
   On the other hand, the contact diagrams at ${\cal O}(q^3)$ and ${\cal O}(q^4)$ enter analytically and the
corresponding LECs can be changed arbitrarily (see Ref.~\cite{Hilt:2013coa} for a discussion of our present values).
   As a specific example, Fig.~\ref{fig:Multipoles} shows the settings to calculate the electric
dipole amplitude $E_{0+}$ for the physical channels at the real-photon point as a function
of the total c.m.~energy $W$.
   The corresponding output is shown in Fig.~\ref{fig:Output}.
   The LECs of the contact interactions can be modified by the user
(see Fig.~\ref{fig:Change_of_parameters}).
   The default settings originate from our fit to the available data
(as at year 2013, see Ref.~\cite{Hilt:2013coa}).

   Of course,  $\chi$MAID has a limited range of applicability.
   First of all, ChPT without additional dynamical degrees of freedom restricts the
energy region, where our results can be applied.
   In the case of neutral pion photoproduction (see Fig.~\ref{fig:wq} and Ref.~\cite{Hilt:2013uf}) one can clearly
see that for energies above $E_\gamma^\textnormal{lab}\approx 170$ MeV the theory starts to deviate
from experimental data.
   The inclusion of the $\Delta(1232)$ resonance at ${\cal O}(q^3)$ has recently been discussed in
Refs.~\cite{Blin:2014rpa,Blin:2015lpa}.
   In the case of the charged channels the range of applicability is larger, but some observables are
quite sensitive to the cutoff of multipoles, as the pion pole term is important at small angles.
   As an estimate, for $W>1160$ MeV the difference between our full amplitude and the approximation
up to and including $G$ waves becomes visible.

\section{Results and conclusions}
\label{sec:results}

\begin{figure}[htbp]
    \centering
        \includegraphics[width=\textwidth]{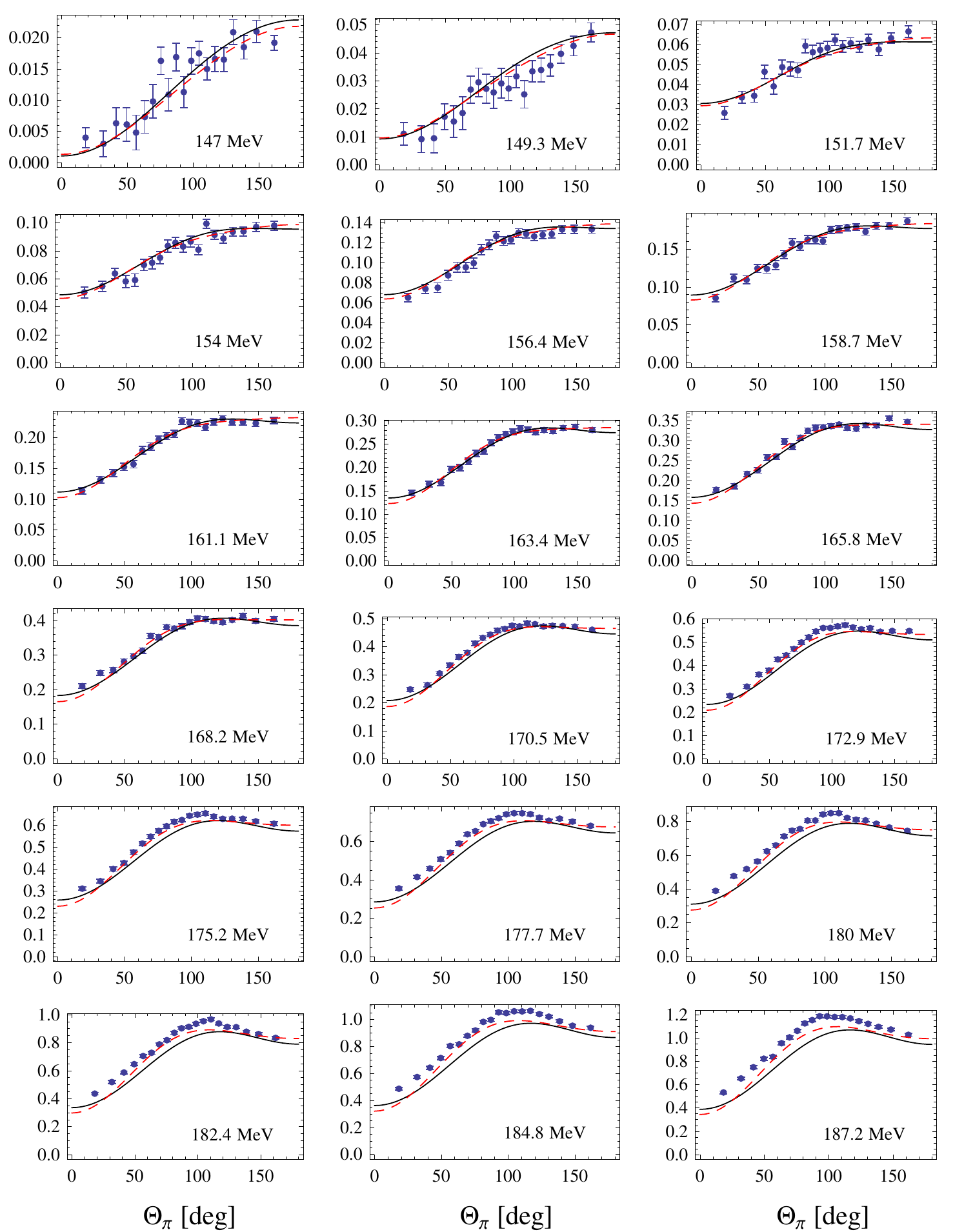}
    \caption{
    Differential cross sections $\sigma_0(\Theta_\pi)$ in $\mu b/sr$ as a function of the
    c.m.~production angle $\Theta_\pi$.
    The graphs are shown for increasing photon energies in the lab frame.
    The solid (black) curves show the results in RChPT at ${\cal O}(q^4)$, the dashed (red) curves show the
    same chiral order in HBChPT.
    The fits make use of data up to and including $E_\gamma^\textnormal{lab}=165.8$ MeV, i.e., the first
    nine panels.
    The data are taken from Ref.\ \cite{Hornidge:2012ca}. The energy $E_\gamma^\textnormal{lab}$ is given
    in each panel.
    }
    \label{fig:wq}
\end{figure}

\begin{figure}[t]
    \centering
        \includegraphics[width=0.5\textwidth]{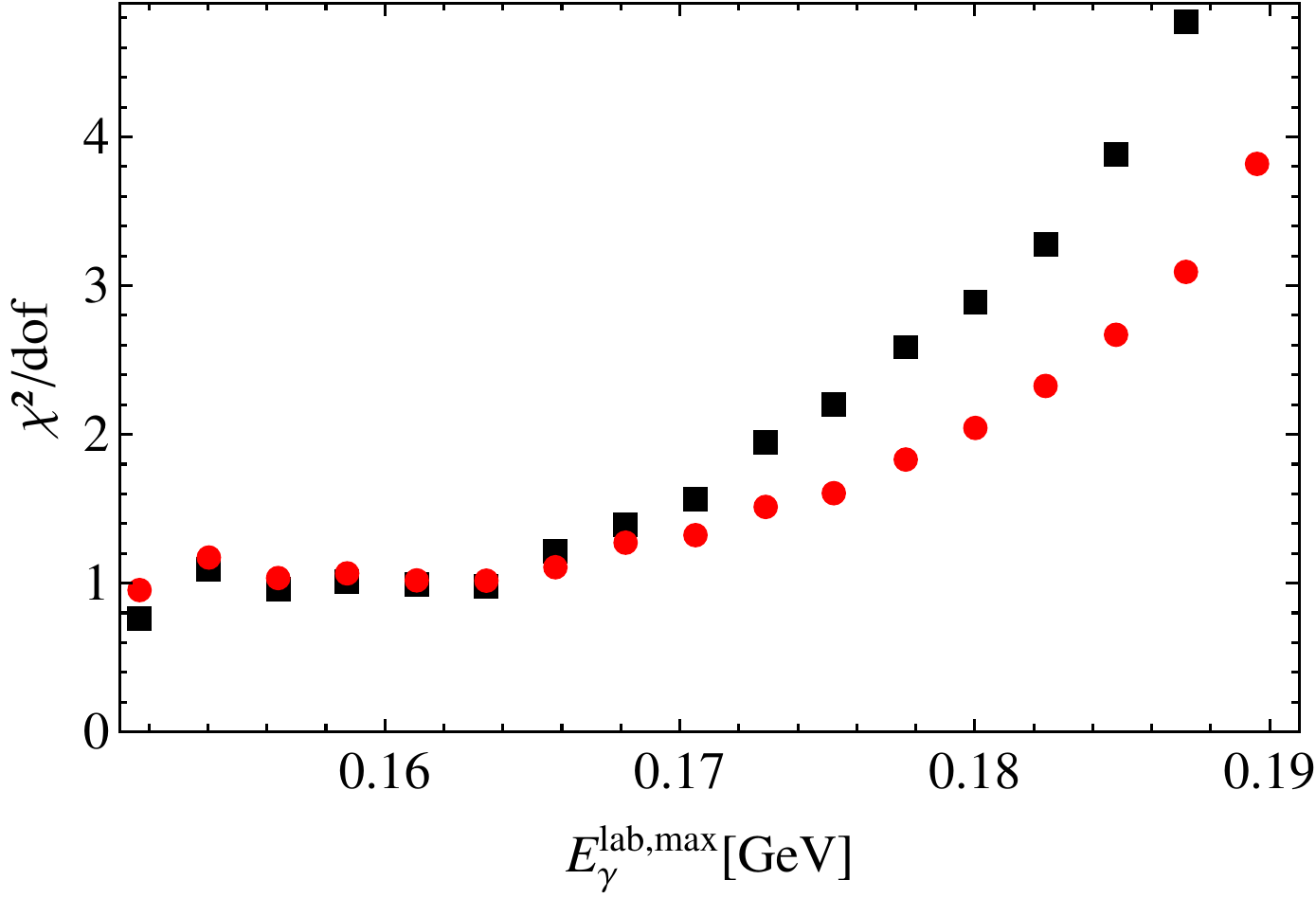}
    \caption{$\chi^2_{\textnormal{red}}$ as a function of the fitted energy range.
    The black squares and red dots refer to the RBChPT and HBChPT fits, respectively.}
    \label{fig:chi2plot}
\end{figure}

\begin{figure}[ht]
    \centering
        \includegraphics[width=\textwidth]{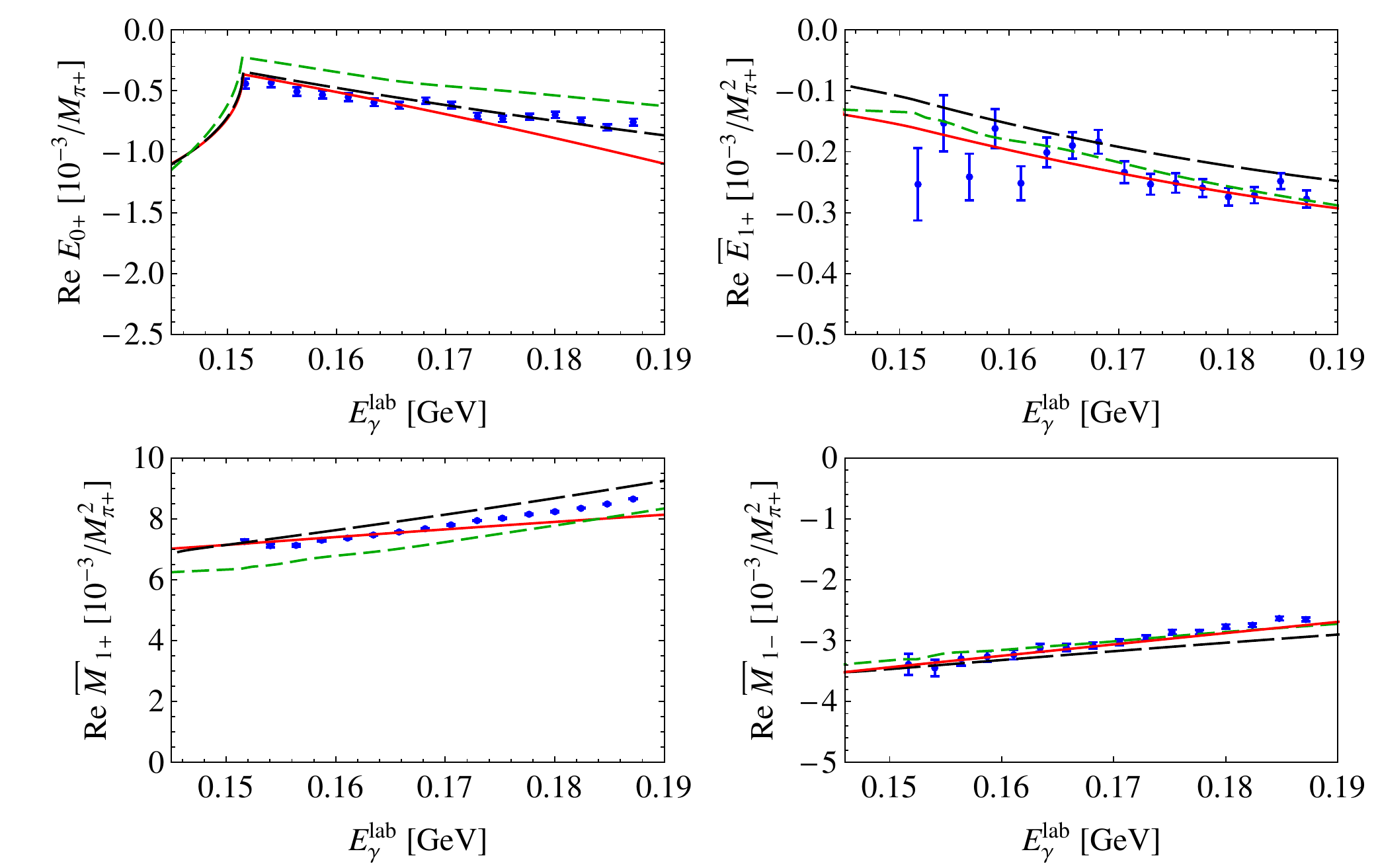}
    \caption{$S$- and reduced $P$-wave multipoles for $\gamma+p\rightarrow p+\pi^0$.
    The solid (red) curves show our RChPT calculations at ${\cal O}(q^4)$.
    The short-dashed (green) and long-dashed (black) curves are
    the predictions of the DMT model \cite{Kamalov:2001qg} and the GL model
    \cite{Gasparyan:2010xz}, respectively.
    The data are from Ref.\ \cite{Hornidge:2012ca}.}
    \label{fig:multipolesexportphyspi0p}
\end{figure}

\begin{figure}[ht]
    \centering
        \includegraphics[width=\textwidth]{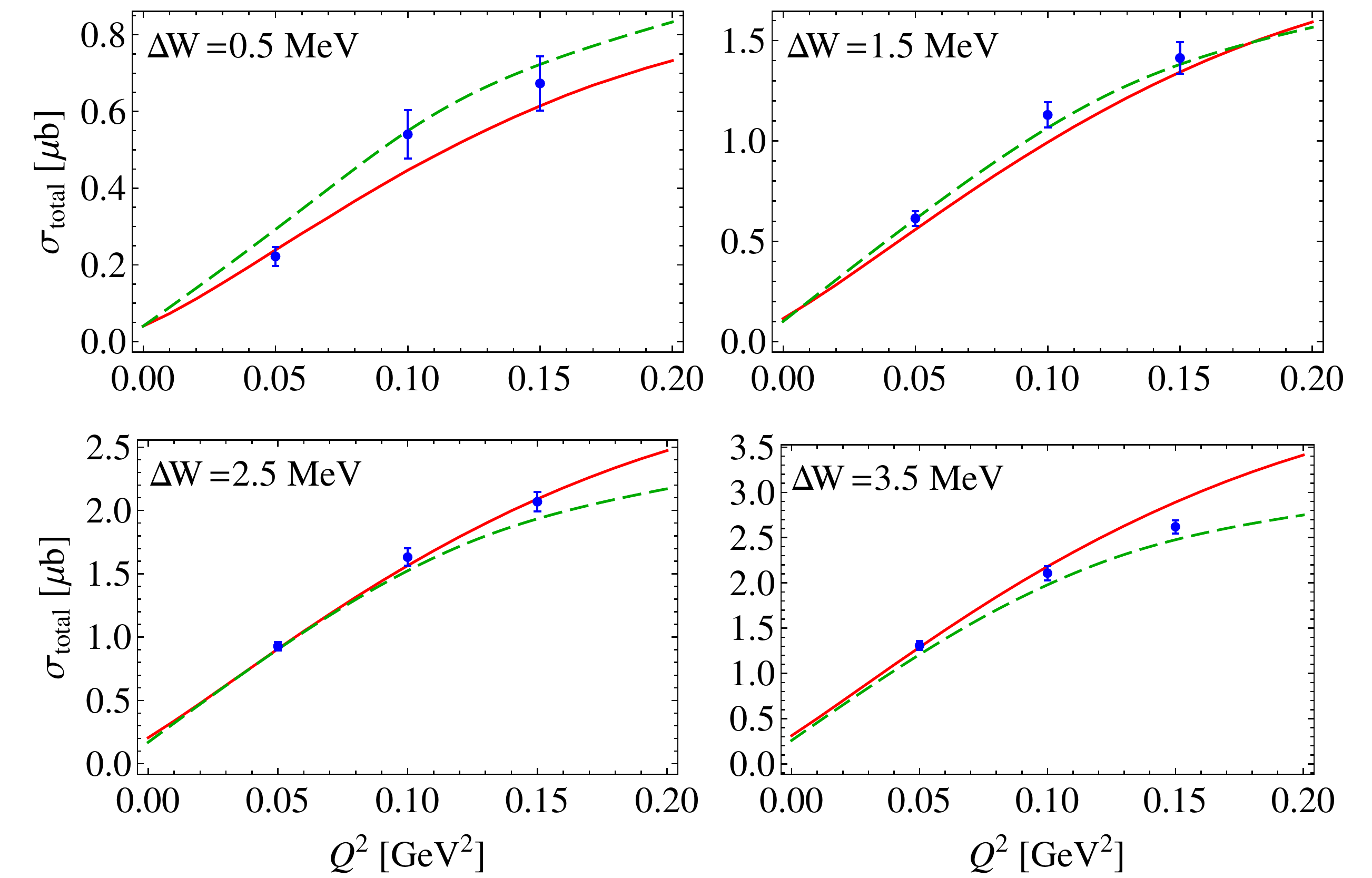}
    \caption{Total cross sections in $\mu$b as a function of $Q^2$ for different c.m.~energies
above threshold $\Delta W$ in MeV. The solid (red) curves show our RChPT calculations at ${\cal O}(q^4)$.
    The short-dashed (green) curves are the predictions of the DMT model \cite{Kamalov:2001qg}.
The data are from Refs.\ \cite{Merkel:20092011}.}
    \label{fig:totcspi0electro}
\end{figure}

\begin{figure}[htbp]
    \centering
        \includegraphics[width=\textwidth]{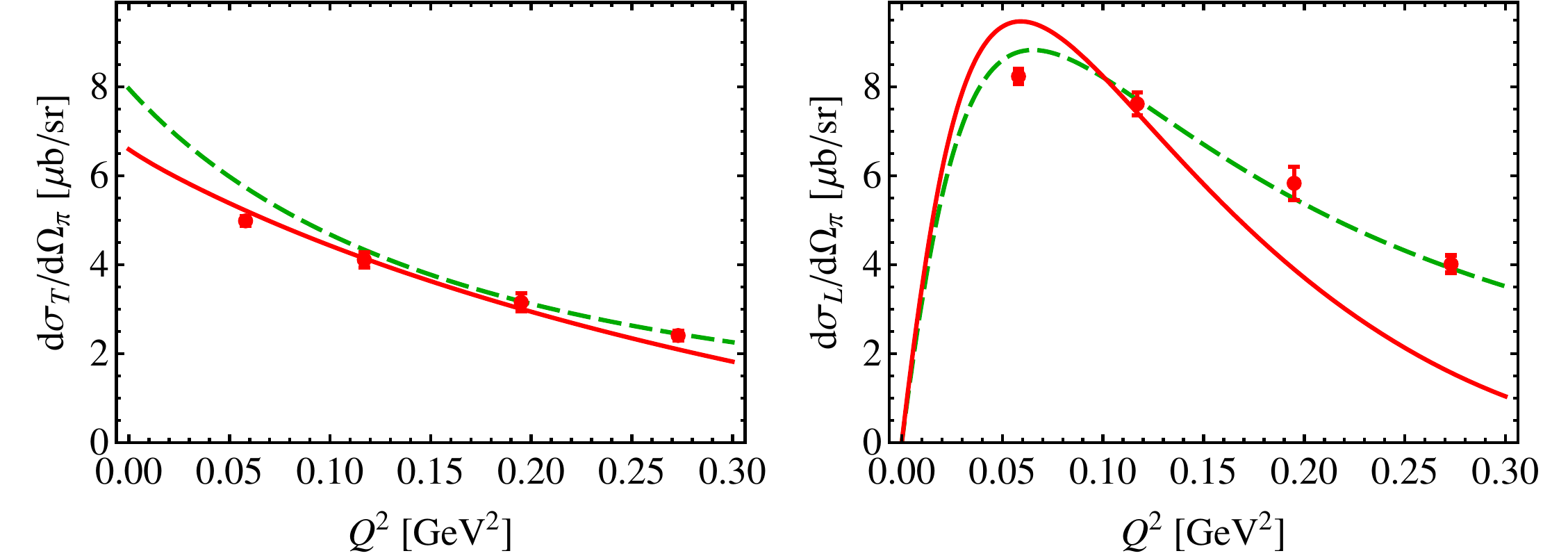}
    \caption{Differential cross sections as a function of $Q^2$ for $\gamma^*+p\rightarrow n+\pi^+$ at
    $W=1125$ MeV and $\Theta_\pi=0^\circ$.
    The solid (red) curves show our RChPT calculation at ${\cal O}(q^4)$ and the dashed (green)
    curves are the predictions of the DMT model \cite{Kamalov:2001qg}.
    The data are from Refs.\ \cite{baumann,Drechsel:2007sq}.}
    \label{fig:chelectro}
\end{figure}

   In the following, we present selected results generated with the chiral MAID
(see Refs.~\cite{Hilt:2013coa,Hilt:2013uf} for a complete discussion).
   First, in Fig.~\ref{fig:wq} we show the differential cross sections $\sigma_0(\Theta_\pi)$
of $\gamma+p\to p+\pi^0$ in $\mu b/sr$ as a function of the c.m.~production angle $\Theta_\pi$.
   The fits make use of data up to and including $E_\gamma^\textnormal{lab}=165.8$ MeV, i.e., the first
nine panels.
   For larger values of $E_\gamma^\textnormal{lab}$, differences between experiment and our calculation
become visible.
   In Fig.\ \ref{fig:chi2plot}, we show how the reduced $\chi^2_\textnormal{red}$ changes if one includes all
data points up to a certain energy $E_\gamma^\textnormal{lab,max}$.
   For comparison we also provide the reduced $\chi^2_\textnormal{red}$ of the HBChPT fit.
   Next, in Fig.\ \ref{fig:multipolesexportphyspi0p} we show the real parts of the $S$ and $P$ waves
of $\gamma+p\to p+\pi^0$ together with single-energy fits of Ref.\ \cite{Hornidge:2012ca}.
   For comparison, we also show the
predictions of the Dubna-Mainz-Taipei (DMT) model \cite{Kamalov:2001qg} and the covariant, unitary,
chiral approach of Gasparyan and Lutz (GL) \cite{Gasparyan:2010xz}. The multipole $E_{0+}$ agrees
nicely with the data in the fitted energy range. The reduced $P$ waves
$\overline{E}_{1+}=E_{1+}/q_\pi$ and $\overline{M}_{1-}=M_{1-}/q_\pi$ with the pion momentum
$q_\pi$ in the c.m.~frame agree for even higher energies with the single energy fits. The largest
deviation can be seen in $\overline{M}_{1+}$. This multipole is related to the $\Delta(1232)$ resonance
and the rising of the data above 170 MeV can be traced back to the influence of this resonance. As
we did not include the $\Delta(1232)$ explicitly, this calculation is not able to fully describe its
impact on the multipole.
   For electroproduction, $\gamma^\ast+p\to p+\pi^0$, in Fig.\ \ref{fig:totcspi0electro}, we show the total cross section
$\sigma_{\rm total}=\sigma_T+\epsilon\sigma_L$ in the threshold region together with the experimental
data \cite{Merkel:20092011}.
   As a final example, we consider the reaction $\gamma^\ast+p\to n+\pi^+$.
   In this reaction channel, only a few data points exist in the energy
range and for photon virtualities, where ChPT can be applied.
   Unfortunately, these data of the differential cross  sections
$\sigma_T$ and $\sigma_L$ at $W=1125$ MeV are at one fixed angle,
namely, $\Theta_\pi=0^\circ$ \cite{baumann,Drechsel:2007sq}.
   The results of our calculation are shown in Fig.\ \ref{fig:chelectro}.
   While the theory agrees with the data for $\sigma_T$, for $\sigma_L$ some
deviation is visible.

   In summary we have shown for the first time a chiral perturbation theory approach that can
consistently describe all pion photo- and electroproduction processes in the threshold region equally well.
   By performing fits to the available experimental data, we determined all 19 LECs of the contact graphs
at ${\cal O}(q^3)$ and ${\cal O}(q^4)$ (see Table \ref{tab:lecvalues}).
   Our relativistic chiral perturbation theory calculation is also available online within the MAID
project as chiral MAID under http://www.kph.uni-mainz.de/MAID/.
   It is clear that new experiments will lead to different estimates for the LECs
\cite{Chirapatpimol:2015ftl,Friscic:2015tga}.
   For that reason, we included in $\chi$MAID the possibility of changing the LECs arbitrarily.
   This will help to further study the range of validity and applicability of ChPT in the future.

\begin{table}[htbp]
\centering
\begin{tabular}{ccc}
\hline
\hline
  Isospin channel   & LEC  &      Value       \\
\hline
 0 & $d_9\ [\textnormal{GeV}^{-2}]$               &       $-1.22\pm0.12$            \\
 0 & $e_{48}\ [\textnormal{GeV}^{-3}]$            &       $5.2\pm1.4$            \\
 0 & $e_{49}\ [\textnormal{GeV}^{-3}]$            &       $0.9\pm2.6$            \\
 0 & $e_{50}\ [\textnormal{GeV}^{-3}]$            &       $2.2\pm0.8$            \\
 0 & $e_{51}\ [\textnormal{GeV}^{-3}]$            &       $6.6\pm3.6$            \\
 0 & $e_{52}^*\ [\textnormal{GeV}^{-3}]$          &       $-4.1$            \\
 0 & $e_{53}^*\ [\textnormal{GeV}^{-3}]$          &       $-2.7$            \\
 0 & $e_{112}\ [\textnormal{GeV}^{-3}]$           &       $9.3\pm1.6$            \\
\hline
 + & $d_8\ [\textnormal{GeV}^{-2}]$               &       $-1.09\pm0.12$            \\
 + & $e_{67}\ [\textnormal{GeV}^{-3}]$            &       $-8.3\pm1.5$            \\
 + & $e_{68}\ [\textnormal{GeV}^{-3}]$            &       $-0.9\pm2.6$            \\
 + & $e_{69}\ [\textnormal{GeV}^{-3}]$            &       $-1.0\pm2.2$            \\
 + & $e_{71}\ [\textnormal{GeV}^{-3}]$            &       $-4.4\pm3.7$            \\
 + & $e_{72}^*\ [\textnormal{GeV}^{-3}]$          &       $10.5$            \\
 + & $e_{73}^*\ [\textnormal{GeV}^{-3}]$          &       $2.1$            \\
 + & $e_{113}\ [\textnormal{GeV}^{-3}]$           &       $-13.7\pm2.6$            \\
\hline
 $-$ & $d_{20}\ [\textnormal{GeV}^{-2}]$            &       $4.34\pm0.08$            \\
 $-$ & $d_{21}\ [\textnormal{GeV}^{-2}]$            &       $-3.1\pm0.1$            \\
 $-$ & $e_{70}\ [\textnormal{GeV}^{-3}]$            &       $3.9\pm0.3$            \\
\hline
\hline
\end{tabular}
\caption{Numerical values of all LECs of pion photo- and electroproduction.
The $\ast$ indicates constants that appear in electroproduction, only.
If possible, the errors were estimated using the bootstrap method (see Ref.~\cite{Hilt:2013coa} for details).
   In the case of the electroproduction LECs $e_{52}$, $e_{53}$, $e_{72}$, and $e_{73}$, we can only give
errors for $\tilde{e}_{52}=e_{52}+e_{72}=6.4\pm 0.7$ and $\tilde{e}_{53}=e_{53}+e_{73}=-0.5\pm 0.2$.}
 \label{tab:lecvalues}
\end{table}

This work was supported by the Deutsche Forschungsgemeinschaft (SFB 443 and 1044).


\begin{thebibliography}{10}
%\cite{Mazzucato:1986dz}
\bibitem{Mazzucato:1986dz}
  E.~Mazzucato  {\it et al.},
  {\it Precise Measurement of Neutral-Pion Photoproduction on the Proton near Threshold},
  Phys.\ Rev.\ Lett.\  {\bf 57} (1986) 3144.
  %%CITATION = PRLTA,57,3144;%%

%\cite{Beck:1990da}
\bibitem{Beck:1990da}
  R.~Beck {\it et al.},
  {\it Measurement of the $p(\gamma,\pi^0)$ Cross Section at Threshold},
  Phys.\ Rev.\ Lett.\  {\bf 65} (1990) 1841.
  %%CITATION = PRLTA,65,1841;%%

%\cite{DeBaenst:1971hp}
\bibitem{DeBaenst:1971hp}
  P.~De Baenst,
  {\it An improvement on the Kroll-Ruderman theorem},
  Nucl.\ Phys.\ {\bf B24} (1970) 633.
  %%CITATION = NUPHA,B24,633;%%

%\cite{Vainshtein:1972ih}
\bibitem{Vainshtein:1972ih}
  A.~I.~Vainshtein and V.~I.~Zakharov,
  {\it Low-energy theorems for photo- and electropion production at threshold},
  Nucl.\ Phys.\ {\bf B36} (1972) 589.
  %%CITATION = NUPHA,B36,589;%%

%\cite{Scherer:1991cy}
\bibitem{Scherer:1991cy}
  S.~Scherer and J.~H.~Koch,
  {\it Pion electroproduction at threshold and PCAC},
  Nucl.\ Phys.\ {\bf A534} (1991) 461.
  %%CITATION = NUPHA,A534,461;%%

%\cite{Bernard:1991rt}
\bibitem{Bernard:1991rt}
  V.~Bernard, N.~Kaiser, J.~Gasser, and U.-G.~Mei\ss ner,
  {\it Neutral pion photoproduction at threshold},
  Phys.\ Lett.\ B {\bf 268} (1991) 291.
  %%CITATION = PHLTA,B268,291;%%

%\cite{Davidson:1993et}
\bibitem{Davidson:1993et}
  R.~M.~Davidson,
  {\it One-loop corrections to threshold $\gamma p \to p \pi^0$ production in the linear sigma model},
  Phys.\ Rev.\ C {\bf 47} (1993) 2492.
  %%CITATION = PHRVA,C47,2492;%%

%\cite{Bernard:1992qa}
\bibitem{Bernard:1992qa}
  V.~Bernard, N.~Kaiser, J.~Kambor, and U.-G.~Mei{\ss}ner,
  {\it Chiral structure of the nucleon},
  Nucl.\ Phys.\ {\bf B388} (1992) 315.
  %%CITATION = NUPHA,B388,315;%%

%\cite{Hilt:2013coa}
\bibitem{Hilt:2013coa}
  M.~Hilt, B.~C.~Lehnhart, S.~Scherer, and L.~Tiator,
  {\it Pion photo- and electroproduction in relativistic baryon chiral perturbation theory and the chiral MAID interface},
  Phys.\ Rev.\ C {\bf 88} (2013) 5, 055207 [arXiv:1309.3385 [nucl-th]].
  %%CITATION = doi:10.1103/PhysRevC.88.055207;%%

%\cite{Becher:1999he}
\bibitem{Becher:1999he}
  T.~Becher and H.~Leutwyler,
  {\it Baryon chiral perturbation theory in manifestly Lorentz invariant form},
  Eur.\ Phys.\ J.\ C {\bf 9} (1999) 643 [hep-ph/9901384].
  %%CITATION = HEP-PH/9901384;%%

%\cite{Gegelia:1999gf}
\bibitem{Gegelia:1999gf}
  J.~Gegelia and G.~Japaridze,
  {\it Matching the heavy particle approach to relativistic theory},
  Phys.\ Rev.\  D {\bf 60} (1999) 114038 [hep-ph/9908377].
  %%CITATION = PHRVA,D60,114038;%%

%\cite{Fuchs:2003qc}
\bibitem{Fuchs:2003qc}
  T.~Fuchs, J.~Gegelia, G.~Japaridze, and S.~Scherer,
  {\it Renormalization of relativistic baryon chiral perturbation theory and power counting},
  Phys.\ Rev.\ D {\bf 68} (2003) 056005 [hep-ph/0302117].
  %%CITATION = HEP-PH/0302117;%%

\bibitem{website}
Chiral MAID, [http://www.kph.uni-mainz.de/MAID/chiralmaid/].

%\cite{Weinberg:1978kz}
\bibitem{Weinberg:1978kz}
  S.~Weinberg,
  {\it Phenomenological Lagrangians},
  Physica A {\bf 96} (1979) 327.
  %%CITATION = PHYSA,A96,327;%%

%\cite{Gasser:1983yg}
\bibitem{Gasser:1983yg}
  J.~Gasser and H.~Leutwyler,
  {\it Chiral perturbation theory to one loop},
  Annals Phys.\  {\bf 158} (1984) 142.
  %%CITATION = APNYA,158,142;%%

%\cite{Gasser:1987rb}
\bibitem{Gasser:1987rb}
  J.~Gasser, M.~E.~Sainio, and A.~\v{S}varc,
  {\it Nucleons with chiral loops},
  Nucl.\ Phys.\ {\bf B307} (1988) 779.
  %%CITATION = NUPHA,B307,779;%%

%\cite{Scherer:2002tk}
\bibitem{Scherer:2002tk}
S.~Scherer, {\it Introduction to Chiral Perturbation Theory},
Adv.\ Nucl.\ Phys.\ {\bf 27} (2003) 277 [hep-ph/0210398].
%%CITATION = ANUPB,27,277;%%

%\cite{Scherer:2012zzd}
\bibitem{Scherer:2012zzd}
  S.~Scherer and M.~R.~Schindler,
  {\it A Primer for Chiral Perturbation Theory},
  Lect.\ Notes Phys.\  {\bf 830} (2012) 1.
  %%CITATION = LNPHA,830,1;%%

%\cite{Collins:xc}
\bibitem{Collins:xc}
J.~C.~Collins, {\em Renormalization}, Cambridge University Press, Cambridge 1984.

%\cite{Weinberg:1995mt}
\bibitem{Weinberg:1995mt}
S.~Weinberg, {\em The Quantum Theory of Fields. Vol.~1: Foundations}, Cambridge University Press, Cambridge
1995.
%\href{http://www.slac.stanford.edu/spires/find/hep/www?irn=3355144}{SPIRES entry}

%\cite{Gegelia:1994zz}
\bibitem{Gegelia:1994zz}
J.~Gegelia, G.~S.~Japaridze, and K.~S.~Turashvili,
{\it Calculation of loop integrals by dimensional counting},
Theor.\ Math.\ Phys.\ {\bf 101} (1994)  1313.
% [Teor.\ Mat.\ Fiz.\  {\bf 101}, 225 (1994)].
%%CITATION = TMFZA,101,225;%%

%\cite{Fuchs:2003sh}
\bibitem{Fuchs:2003sh}
  T.~Fuchs, M.~R.~Schindler, J.~Gegelia, and S.~Scherer,
  {\it Power counting in baryon chiral perturbation theory including vector mesons},
  Phys.\ Lett.\ B {\bf 575} (2003) 11 [hep-ph/0308006].
  %%CITATION = HEP-PH/0308006;%%

%\cite{Hacker:2005fh}
\bibitem{Hacker:2005fh}
  C.~Hacker, N.~Wies, J.~Gegelia, and S.~Scherer,
  {\it Including the $\Delta(1232)$ resonance in baryon chiral perturbation theory},
  Phys.\ Rev.\ C {\bf 72} (2005) 055203 [hep-ph/0505043].
  %%CITATION = HEP-PH/0505043;%%

%\cite{Schindler:2003je}
\bibitem{Schindler:2003je}
  M.~R.~Schindler, J.~Gegelia, and S.~Scherer,
 {\it Infrared and extended on-mass-shell renormalization of two-loop diagrams},
  Nucl.\ Phys.\ {\bf B682} (2004) 367
  [hep-ph/0310207].
  %%CITATION = HEP-PH/0310207;%%

%\cite{Schindler:2003xv}
\bibitem{Schindler:2003xv}
  M.~R.~Schindler, J.~Gegelia, and S.~Scherer,
  {\it Infrared regularization of baryon chiral perturbation theory reformulated},
  Phys.\ Lett.\ B {\bf 586} (2004) 258 [hep-ph/0309005].
  %%CITATION = HEP-PH/0309005;%%

%\cite{Fuchs:2003ir}
\bibitem{Fuchs:2003ir}
  T.~Fuchs, J.~Gegelia, and S.~Scherer,
  {\it Electromagnetic form factors of the nucleon in relativistic baryon chiral perturbation theory},
  J.\ Phys.\ G {\bf 30} (2004) 1407 [nucl-th/0305070].
  %%CITATION = NUCL-TH/0305070;%%
  %39 citations counted in INSPIRE as of 23 Oct 2015

%\cite{Schindler:2005ke}
\bibitem{Schindler:2005ke}
  M.~R.~Schindler, J.~Gegelia, and S.~Scherer,
  {\it Electromagnetic form factors of the nucleon in chiral perturbation theory including vector mesons},
  Eur.\ Phys.\ J.\ A {\bf 26} (2005) 1 [nucl-th/0509005].
  %%CITATION = NUCL-TH/0509005;%%

%\cite{Bauer:2012pv}
\bibitem{Bauer:2012pv}
  T.~Bauer, J.~C.~Bernauer, and S.~Scherer,
  {\it Electromagnetic form factors of the nucleon in effective field theory},
  Phys.\ Rev.\ C {\bf 86} (2012) 065206
  [arXiv:1209.3872 [nucl-th]].
  %%CITATION = ARXIV:1209.3872;%%

%\cite{Schindler:2006it}
\bibitem{Schindler:2006it}
  M.~R.~Schindler, T.~Fuchs, J.~Gegelia, and S.~Scherer,
  {\it Axial, induced pseudoscalar, and pion-nucleon form factors in manifestly Lorentz-invariant chiral perturbation theory},
  Phys.\ Rev.\ C {\bf 75} (2007) 025202 [nucl-th/0611083].
  %%CITATION = NUCL-TH/0611083;%%

%\cite{Alarcon:2011zs}
\bibitem{Alarcon:2011zs}
  J.~M.~Alarc\'on, J.~Martin Camalich, and J.~A.~Oller,
  {\it The chiral representation of the $\pi N$ scattering amplitude and the pion-nucleon sigma term},
  Phys.\ Rev.\ D {\bf 85} (2012) 051503
  [arXiv:1110.3797 [hep-ph]].
  %%CITATION = ARXIV:1110.3797;%%

%\cite{Alarcon:2012kn}
\bibitem{Alarcon:2012kn}
  J.~M.~Alarc\'on, J.~Martin Camalich, and J.~A.~Oller,
  {\it Improved description of the $\pi N$-scattering phenomenology in covariant baryon chiral perturbation theory},
  Annals Phys.\  {\bf 336} (2013) 413
  [arXiv:1210.4450 [hep-ph]].
  %%CITATION = ARXIV:1210.4450;%%

%\cite{Chen:2012nx}
\bibitem{Chen:2012nx}
  Y.~H.~Chen, D.~L.~Yao, and H.~Q.~Zheng,
  {\it Analyses of pion-nucleon elastic scattering amplitudes up to $O(p^4)$ in extended-on-mass-shell subtraction scheme},
  Phys.\ Rev.\ D {\bf 87} (2013) 054019 [arXiv:1212.1893 [hep-ph]].
  %%CITATION = ARXIV:1212.1893;%%

%\cite{Schindler:2006ha}
\bibitem{Schindler:2006ha}
  M.~R.~Schindler, D.~Djukanovic, J.~Gegelia, and S.~Scherer,
  {\it Chiral expansion of the nucleon mass to order  ${\cal O}(q^6)$},
  Phys.\ Lett.\ B {\bf 649} (2007) 390 [hep-ph/0612164].
  %%CITATION = HEP-PH/0612164;%%

%\cite{Schindler:2007dr}
\bibitem{Schindler:2007dr}
  M.~R.~Schindler, D.~Djukanovic, J.~Gegelia, and S.~Scherer,
  {\it Infrared renormalization of two-loop integrals and the chiral expansion of the nucleon mass},
  Nucl.\ Phys.\ {\bf A803} (2008) 68 [arXiv:0707.4296 [hep-ph]].
  %%CITATION = ARXIV:0707.4296;%%

%\cite{Geng:2013xn}
\bibitem{Geng:2013xn}
  L.~Geng,
  {\it Recent developments in SU(3) covariant baryon chiral perturbation theory},
  Front.\ Phys.\ China {\bf 8} (2013) 328 [arXiv:1301.6815 [nucl-th]].
  %%CITATION = ARXIV:1301.6815;%%

%\cite{Djukanovic:2009zn}
\bibitem{Djukanovic:2009zn}
  D.~Djukanovic, J.~Gegelia, A.~Keller, and S.~Scherer,
  {\it Complex-mass renormalization in chiral effective field theory},
  Phys.\ Lett.\ B {\bf 680} (2009) 235
  [arXiv:0902.4347 [hep-ph]].
  %%CITATION = ARXIV:0902.4347;%%

%\cite{Stuart:1990}
\bibitem{Stuart:1990}
R.~G.~Stuart, {\it Pitfalls of radiative corrections near a resonance},
in ${\rm Z}^0$ {\it Physics}, ed.\ by J.~Tran Thanh Van
(Editions Fronti\`eres, Gif-sur-Yvette, 1990) p.\ 41.
  %%CITATION = INSPIRE-306680;%%


%\cite{Denner:1999gp}
\bibitem{Denner:1999gp}
  A.~Denner, S.~Dittmaier, M.~Roth, and D.~Wackeroth,
  {\it Predictions for all processes $e^+ e^-\to fermions + \gamma$},
  Nucl.\ Phys.\  {\bf B560} (1999) 33  [arXiv:hep-ph/9904472].
  %%CITATION = NUPHA,B560,33;%%


%\cite{Djukanovic:2009gt}
\bibitem{Djukanovic:2009gt}
  D.~Djukanovic, J.~Gegelia, and S.~Scherer,
  {\it Chiral structure of the Roper resonance using complex-mass scheme},
  Phys.\ Lett.\ B {\bf 690} (2010) 123
  [arXiv:0903.0736 [hep-ph]].
  %%CITATION = ARXIV:0903.0736;%%

%\cite{Bauer:2012at}
\bibitem{Bauer:2012at}
  T.~Bauer, J.~Gegelia, and S.~Scherer,
  {\it Magnetic moment of the Roper resonance},
  Phys.\ Lett.\ B {\bf 715} (2014) 234
  [arXiv:1208.2598 [hep-ph]].
  %%CITATION = ARXIV:1208.2598;%%

%\cite{Djukanovic:2013mka}
\bibitem{Djukanovic:2013mka}
  D.~Djukanovic, E.~Epelbaum, J.~Gegelia, and U.-G.~Mei{\ss}ner,
  {\it The magnetic moment of the $\rho$-meson},
  Phys.\ Lett.\ B {\bf 730} (2014) 115
  [arXiv:1309.3991 [hep-ph]].
  %%CITATION = ARXIV:1309.3991;%%

%\cite{Bauer:2014cqa}
\bibitem{Bauer:2014cqa}
  T.~Bauer, S.~Scherer, and L.~Tiator,
  {\it Electromagnetic transition form factors of the Roper resonance in a phenomenological field theory},
  Phys.\ Rev.\ C {\bf 90} (2014) 1,  015201
  [arXiv:1402.0741 [nucl-th]].
  %%CITATION = ARXIV:1402.0741;%%

%\cite{Djukanovic:2014rua}
\bibitem{Djukanovic:2014rua}
  D.~Djukanovic, J.~Gegelia, A.~Keller, S.~Scherer, and L.~Tiator,
  {\it Vector form factor of the pion in chiral effective field theory},
  Phys.\ Lett.\ B {\bf 742} (2015) 55
  [arXiv:1410.3801 [hep-ph]].
  %%CITATION = ARXIV:1410.3801;%%

%\cite{Epelbaum:2015vea}
\bibitem{Epelbaum:2015vea}
  E.~Epelbaum, J.~Gegelia, U.~G.~Mei{\ss}ner, and D.~L.~Yao,
  {\it Baryon chiral perturbation theory extended beyond the low-energy region},
  Eur.\ Phys.\ J.\ C {\bf 75} (2015) 499 [arXiv:1510.02388 [hep-ph]].
  %%CITATION = doi:10.1140/epjc/s10052-015-3728-7;%%

%\cite{Chew:1957tf}
\bibitem{Chew:1957tf}
  G.~F.~Chew, M.~L.~Goldberger, F.~E.~Low, and Y.~Nambu,
  {\it Relativistic Dispersion Relation Approach to Photomeson Production},
  Phys.\ Rev.\  {\bf 106} (1957) 1345.
  %%CITATION = PHRVA,106,1345;%%

%\cite{Dennery:1961zz}
\bibitem{Dennery:1961zz}
  P.~Dennery,
  {\it Theory of the Electro- and Photoproduction of $\pi$ Mesons},
  Phys.\ Rev.\  {\bf 124} (1961) 2000.
  %%CITATION = PHRVA,124,2000;%%

%\cite{Amaldi:1979vh}
\bibitem{Amaldi:1979vh}
  E.~Amaldi, S.~Fubini, and G.~Furlan,
  {\it Pion-Electroproduction. Electroproduction at Low Energy and Hadron Form Factors},
  Springer Tracts Mod.\ Phys.\ {\bf 83} (1979) 1.
  %%CITATION = STPHB,83,1;%%


%\cite{Hornidge:2012ca}
\bibitem{Hornidge:2012ca}
  D.~Hornidge {\it et al.}  [A2 and CB-TAPS Collaboration],
  {\it Accurate Test of Chiral Dynamics in the $\vec{\gamma} p \to \pi^0p$ Reaction},
  Phys.\ Rev.\ Lett.\  {\bf 111} (2013) 062004  [arXiv:1211.5495 [nucl-ex]].
  %%CITATION = ARXIV:1211.5495;%%

%\cite{Drechsel:1992pn}
\bibitem{Drechsel:1992pn}
  D.~Drechsel and L.~Tiator,
  {\it Threshold pion photoproduction on nucleons},
  J.\ Phys.\ G {\bf 18} (1992) 449.
  %%CITATION = JPHGB,G18,449;%%

%\cite{Mertig:1990an}
\bibitem{Mertig:1990an}
  R.~Mertig, M.~Bohm, and A.~Denner,
  {\it Feyn Calc -- Computer-algebraic calculation of Feynman amplitudes},
  Comput.\ Phys.\ Commun.\  {\bf 64} (1991) 345.
  %%CITATION = CPHCB,64,345;%%

%\cite{Hahn:2000kx}
\bibitem{Hahn:2000kx}
  T.~Hahn,
  {\it Generating Feynman diagrams and amplitudes with FeynArts 3},
  Comput.\ Phys.\ Commun.\  {\bf 140} (2001) 418 [hep-ph/0012260].
  %%CITATION = HEP-PH/0012260;%%

%\cite{Hilt:2013uf}
\bibitem{Hilt:2013uf}
  M.~Hilt, S.~Scherer, and L.~Tiator,
  {\it Threshold $\pi^0$ photoproduction in relativistic chiral perturbation theory},
  Phys.\ Rev.\ C {\bf 87} (2013) 045204 [arXiv:1301.5576 [nucl-th]].
  %%CITATION = ARXIV:1301.5576;%%

%\cite{Blin:2014rpa}
\bibitem{Blin:2014rpa}
  A.~N.~Hiller Blin, T.~Ledwig, and M.~J.~Vicente Vacas,
  {\it Chiral dynamics in the $\vec{\gamma}p \to p\pi^0$ reaction},
  Phys.\ Lett.\ B {\bf 747} (2015) 217 [arXiv:1412.4083 [hep-ph]].
  %%CITATION = ARXIV:1412.4083;%%

%\cite{Blin:2015lpa}
\bibitem{Blin:2015lpa}
  A.~H.~Blin, T.~Ledwig, and M.~V.~Vacas,
  {\it Neutral pion photoproduction on protons in fully covariant ChPT with $\Delta(1232)$ loop contributions},
  arXiv:1510.01598 [hep-ph].
  %%CITATION = ARXIV:1510.01598;%%

%\cite{Kamalov:2001qg}
\bibitem{Kamalov:2001qg}
  S.~S.~Kamalov, G.-Y.~Chen, S.-N.~Yang, D.~Drechsel, and L.~Tiator,
  {\it $\pi^0$ Photo- and electroproduction at threshold within a dynamical model},
  Phys.\ Lett.\ B {\bf 522} (2001) 27 [nucl-th/0107017].
  %%CITATION = NUCL-TH/0107017;%%

%\cite{Gasparyan:2010xz}
\bibitem{Gasparyan:2010xz}
  A.~Gasparyan and M.~F.~M.~Lutz,
  {\it Photon- and pion-nucleon interactions in a unitary and causal effective field theory based on the chiral Lagrangian},
  Nucl.\ Phys.\ {\bf A848} (2010) 126 [arXiv:1003.3426 [hep-ph]].
  %%CITATION = ARXIV:1003.3426;%%

\bibitem{Merkel:20092011}
  H.~Merkel, {\it Experimental results from MAMI}, PoS CD {\bf 09} (2009) 112;
  H.~Merkel {\it et al.}, {\it Consistent threshold $\pi^0$ electro-production at $Q^2=0.05$, $0.10$, and 0.15 GeV$^2$/$c^2$}, arXiv:1109.5075 [nucl-ex].
  %%CITATION = POSCI,CD09,112;%%
  %%CITATION = ARXIV:1109.5075;%%

%\cite{Weis:2007kf}
\bibitem{Weis:2007kf}
  M.~Weis {\it et al.}  [A1 Collaboration],
  {\it Separated cross-sections in $\pi^0$ electroproduction at threshold at $Q^2=0.05$ GeV$^2$/$c^2$},
  Eur.\ Phys.\ J.\ A {\bf 38} (2008) 27 [arXiv:0705.3816 [nucl-ex]].
  %%CITATION = ARXIV:0705.3816;%%

\bibitem{baumann}
D.~Baumann, {\it $\pi^+$-Elektroproduktion an der Schwelle}, PhD thesis (in German),
%Johannes Gutenberg-Universit\"at Mainz,
JGU Mainz,
2005, [http://ubm.opus.hbz-nrw.de/volltexte/2006/923/pdf/diss.pdf].

%\cite{Drechsel:2007sq}
\bibitem{Drechsel:2007sq}
  D.~Drechsel and T.~Walcher,
 {\it Hadron structure at low $Q^2$},
  Rev.\ Mod.\ Phys.\  {\bf 80} (2008) 731
  [arXiv:0711.3396 [hep-ph]].
  %%CITATION = ARXIV:0711.3396;%%

%\cite{Chirapatpimol:2015ftl}
\bibitem{Chirapatpimol:2015ftl}
  K.~Chirapatpimol {\it et al.} [Hall A Collaboration],
  {\it Precision Measurement of the $p(e,e'p)\pi^0$ Reaction at Threshold},
  Phys.\ Rev.\ Lett.\  {\bf 114} (2015) 19, 192503 [arXiv:1501.05607 [nucl-ex]].
  %%CITATION = ARXIV:1501.05607;%%

%\cite{Friscic:2015tga}
\bibitem{Friscic:2015tga}
  I.~Fri\v{s}\v{c}i\'c, {\it Measurement of the $p(e,e'\pi^+)n$ reaction with the short-orbit spectrometer at $Q^2 = 0.078$ (GeV/c)$^{2}$},
  PhD thesis, University of Zagreb, 2015.
  %%CITATION = INSPIRE-1358259;%%

\end{thebibliography}
\end{document}